\documentclass[pdflatex,sn-mathphys-num]{sn-jnl}

\usepackage{graphicx}%
\usepackage{multirow}%
\usepackage{amsmath,amssymb,amsfonts}%
\usepackage{amsthm}%
\usepackage{mathrsfs}%
\usepackage[title]{appendix}%
\usepackage{xcolor}%
\usepackage{textcomp}%
\usepackage{manyfoot}%
\usepackage{booktabs}%
\usepackage{algorithm}%
\usepackage{algorithmicx}%
\usepackage{algpseudocode}%
\usepackage{listings}%
\usepackage{mathtools}

\usepackage{color,xcolor,ucs}
\usepackage{mathtools}   \usepackage{tikz} 
\usepackage{ amssymb }
\usepackage{extarrows} 
\usepackage{pgf,tikz}
\usepackage{float}
\usetikzlibrary{positioning}
\usetikzlibrary{shapes.geometric}
\usetikzlibrary{shapes.misc}
\usetikzlibrary{arrows}
\usepackage{caption}
\usepackage{mathrsfs}
\usetikzlibrary{arrows,shapes,automata,backgrounds,petri,positioning}
\usetikzlibrary{decorations.pathmorphing}
\usetikzlibrary{decorations.shapes}
\usetikzlibrary{decorations.text}
\usetikzlibrary{decorations.fractals}
\usetikzlibrary{decorations.footprints}
\usetikzlibrary{shadows}
\usetikzlibrary{calc}
\usetikzlibrary{spy}
\usepackage{amsmath}
\usepackage{array}
\usepackage{ amssymb }
\usepackage{braket}
\usepackage{qcircuit}
\usepackage{soul}
\usepackage{braket} 
\usepackage{relsize}

\usepackage{amsmath}
\usepackage{ amssymb }
\usepackage{braket}
\usepackage{qcircuit}
\usepackage{soul}
\usepackage{braket}

\theoremstyle{thmstyleone}%
\theoremstyle{thmstyletwo}%

\theoremstyle{thmstylethree}%

\begin{document}

\title[Article Title]{Intertwining vectors, and Boltzmann weight matrices, of a Solid-on-Solid model from the 20-vertex model}


\author[1]{\fnm{Pete} \sur{Rigas}}\email{pbr43@cornell.edu}


\abstract{We initiate a new study on the correspondence between the 20-vertex model and a SOS (Solid-on-Solid) model. In comparison to two previous works of the author in 2024 which characterized properties of the transfer, and quantum monodromy, matrices of the 20-vertex model from the perspective of the quantum inverse scattering method, in addition to the structure of nonlocal correlations, the forthcoming approach is an adaptation of a study on the rational 7-vertex model. For the rational 7-vertex model,  Antoenneko and Valinevich demonstrated that the intertwining vectors can be used to transform the R-matrix of a vertex model into the Boltzman weight matrix of an SOS model. To further develop perspectives on classes of higher dimensional SOS models, by leveraging previous computations with L-operators due to the author, and by also manipulating q-exponentials, and other related factors from a factorization of the universal R-matrix due to Boos et al, analogs of intertwining vectors for the 20-vertex model can be obtained. In comparison to the system of equations that have been obtained for the rational 7-vertex model, the system that one obtains for the 20-vertex model consists of 9 equations for each entry of the R-matrix. Furthermore, from the fact that the 20-vertex model contains more possible configurations in the sample space than the rational 7-vertex model does, the Boltzman weight matrix of the corresponding SOS model asymptotically depends upon several contributions, a few of which include tensor products of basis elements, and spectral parameters over the triangular lattice.  }

\keywords{Statistical Physics, Mathematical Physics, eigenvalue problem, SOS model, 6-vertex model, 7-vertex model, 20-vertex model, 4-vertex model}


\pacs[MSC Classification]{34L25, 60K35}

\maketitle

\section{Introduction}\label{sec1}

\subsection{Integrable Probabilistic objects}

Vertex models of Statistical Physics have long attracted attention from a wide variety of perspectives, ranging from exact solvability, {\color{blue}[3},{\color{blue}5]}, constructions of universal R-matrices, {\color{blue}[6},{\color{blue}7]}, anti-periodicity in boundary conditions, {\color{blue}[4]}, classes of spin-boson models from rational vertex models, {\color{blue}[2]}, the arctic circle {\color{blue}[9]}, amongst several other connections between Mathematical and Statistical Physics {\color{blue}[8},{\color{blue}10},{\color{blue}11},{\color{blue}12},{\color{blue}13},{\color{blue}14},{\color{blue}15},{\color{blue}17},{\color{blue}18},{\color{blue}19},{\color{blue}20]}. In previous works of the author, classes of properties that are similar to integrability properties of the 6-vertex model, can be explored for the 20-vertex model through asymptotic representations of the quantum monodromy matrix,

\begin{align*}
 T^{3D}_{a,b} \big(    \big\{ u_i \big\} , \big\{ v^{\prime}_j  \big\} , \big\{ w^{\prime\prime}_k \big\}     \big) :   \textbf{C}^3 \otimes \big( \textbf{C}^3 \big)^{\otimes ( |N| + ||M||_1 )}  \longrightarrow   \textbf{C}^3 \otimes \big( \textbf{C}^3 \big)^{\otimes ( |N| + ||M||_1 )} \\   \mapsto  \overset{-N}{\underset{j=0}{\prod}} \text{ }  \overset{\underline{M}}{\underset{k=0}{\prod}} \bigg[  \mathrm{diag} \big( \mathrm{exp} \big(  \alpha \big(i, j,k \big)  \big)   , \mathrm{exp} \big(  \alpha \big( i,j,k\big)  \big)   , \mathrm{exp} \big(  \alpha \big( i, j,k \big)  \big) \big)  R_{ia,jb,kc} \big( u - u_i ,  u^{\prime} - v^{\prime}_j , w-w^{\prime\prime}_k \big) \bigg]  \text{, } 
\end{align*}

\noindent where $N,M$ denote two naturals, $\alpha \big( i, j, k \big)$ denotes some sufficiently smooth function, while the $R$ term denotes the Universal R-matrix, which has the factorization,

\[
R = R_{\leq \delta} R_{\sim \delta} R_{\geq \delta} K \text{, }
\]

\noindent for,

\begin{align*}
 R_{\leq \delta} \equiv   \underset{m \in \textbf{N}}{\underset{\gamma \in \Delta_{+} ( A)}{\prod}}       \mathrm{exp}_q    \bigg[   \big( q - q^{-1} \big)            s^{-1}_{m , \gamma} e_{\gamma + m \delta} \otimes f_{\gamma + m \delta}          \bigg] \text{, } \end{align*}
 
 \begin{align*}
 R_{\sim \delta}  \equiv     \mathrm{exp} \bigg[  \big( q - q^{-1} \big) \underset{m \in \textbf{Z}^{+}}{\sum}  \text{ }   \overset{r}{\underset{i , j =1}{\sum}}      u_{m,ij} e_{m \delta , \alpha_i } \otimes f_{m \delta , \alpha_j }        \bigg]    \text{, } \\ 
 R_{\geq \delta} \equiv           \underset{m \in \textbf{N}}{\underset{\gamma \in \Delta_{+} ( A)}{\prod}}       \mathrm{exp}_q    \bigg[   \big( q - q^{-1} \big)            s^{-1}_{m , \delta - \gamma} e_{\delta - \gamma + m \delta} \otimes f_{ \delta - \gamma + m \delta}          \bigg]      \text{, } \\  K \equiv   \mathrm{exp} \bigg[   \hbar     \overset{r}{ \underset{i , j =1 }{\sum} } \big(  b_{ij} h_{\alpha_i} \otimes h_{\alpha_j} \big)   \bigg]    \text{. } 
\end{align*}

\noindent The K-matrix, from the last component of the universal R-Matrix above, consists of a scaling dependent upon the Planck constant, while the q-exponential factors appearing from other contributions of the factorization take the form,

\begin{align*}
   \mathrm{exp}_q \big[ z \big] \equiv  \overset{+ \infty}{\underset{n=0}{\sum}}      \frac{z^n}{\big[ n \big]_q !}      \equiv      \overset{+ \infty}{\underset{n=0}{\sum}}   z^n  \bigg[ \frac{\big( 1 - q \big)^n}{\underset{1 \leq i \leq n}{\prod} \big( 1 - q^i \big)      }   \bigg]              \text{. }
\end{align*}

\noindent In the presence of two external fields, one of which is taken to be vanishing, the mapping above over tensor products in $\textbf{C}^3$ takes the form,

\begin{align*}
 T^{3D}_{a,b} \big(     \big\{ u_i \big\} , \big\{ v^{\prime}_j  \big\} , \big\{ w^{\prime\prime}_k \big\}   , H , 0     \big) :   \textbf{C}^3 \otimes \big( \textbf{C}^3 \big)^{\otimes ( |N| + ||M||_1 )}  \longrightarrow   \textbf{C}^3 \otimes \big( \textbf{C}^3 \big)^{\otimes ( |N| + ||M||_1 )} \\   \mapsto  \overset{-N}{\underset{j=0}{\prod}} \text{ }  \overset{\underline{M}}{\underset{k=0}{\prod}} \bigg[  \mathrm{diag} \big( \mathrm{exp} \big( H \big)   , \mathrm{exp} \big(  H\big)  , \mathrm{exp} \big( H \big) \big)  R_{ia,jb,kc} \big( u - u_i ,  u^{\prime} - v^{\prime}_j , w-w^{\prime\prime}_k \big) \bigg]  \text{. } 
\end{align*}

\noindent Each one of the expressions above for the quantum monodromy matrix can be related to the transfer matrix, which takes the form,

\begin{align*}
 \textbf{T}^{3D} \big( \underline{\lambda} \big)   \equiv    \underset{ N \longrightarrow -  \infty}{\underset{\underline{M} \longrightarrow + \infty}{\mathrm{lim}}} \mathrm{tr} \bigg[     \overset{\underline{M}}{\underset{j=0}{\prod}}  \text{ }  \overset{-N}{\underset{k=0}{\prod}} \mathrm{exp} \big( \lambda_3 ( q^{-2} \xi^{s^j_k} ) \big)     \bigg[ \begin{smallmatrix}     q^{D^j_k}       &    q^{-2} a^j_k q^{-D^j_k -D^j_{k+1}} \xi^{s-s^k_j}        &  a^j_k a^j_{k+1} q^{-D^j_k - 3D^j_{k+1}} \xi^{s - s^j_k - s^j_{k+1}}  \\ \big( a^j_k \big)^{\dagger} q^{D^j_k} \xi^{s^j_k} 
             &   *_2   &     - a^j_k q^{D^j_k - 3D^j_{k+1}} \xi^{s-s^j_k}  \\ 0  &    a^{\dagger}_j q^{D^j_k} \xi^{s^j_k} &  q^{-D^j_k} \\    \end{smallmatrix} \bigg]      \bigg] 
\text{, }
\end{align*}

\noindent where,

\begin{align*}
    *_2 \equiv    q^{-D^j_k + D^j_{k+1}} - q^{-2} q^{D^j_k -D^j_{k+1}} \xi^{s}   \text{, }
\end{align*}

\noindent in which the product over $\underline{M}$, and $N$, is dependent upon three-dimensional L-operators introduced from the factorization of the Universal R-matrix in {\color{blue}[7]}, for the spectral parameter $\underline{\lambda}$ {\color{blue}[45]}. With the $q$-parameters introduced in the three-dimensional L-operator above, denote the differential operators with,

\begin{align*}
D^j_k  \equiv \big(  D      \otimes \textbf{1} \big) \textbf{1}_{\{\textbf{r} \equiv e_k\}}  \text{, } \\  D^j_{k+1}\equiv \big(  \textbf{1} \otimes D \big) \textbf{1}_{\{\textbf{r} \equiv e_{k+1}\}}  \text{, }  
\end{align*}

 With such objects in place, to establish a correspondence between the Universal R-matrix of the 20-vertex model, and the Boltzmann weight matrix of a SOS model, we make use of several observations previously developed for the rational 7-vertex SOS model {\color{blue}[1]}. In such computations for the 7-vertex model, the authors make significant use of an intertwining relation, which is used to transform the R-matrix of the 7-vertex model into the Boltzmann weight matrix of the 7-vertex SOS model.

To determine whether other characteristics of interest exist for the 20-vertex SOS model, one must follow a program consisting of several steps, the first of which consists in beginning with the intertwining relation for the 20-vertex model, insofar as being able to compute the intertwining vectors; a decomposition, from the factorization of the Universal R-matrix provided above from {\color{blue}[7]}, into contributions from combinatorial q-exponential factors, and K-matrices; and finally, a derivation of a system of nine equations from which the form of the intertwining vectors can be read off. In being able to convert a problem of intertwining vectors which act on the Universal R-matrix to a system of relations, we essentially focus our study of the 20-vertex SOS model to an eigenvalue problem, which has been previously examined by the author for Toeplitz matrices, Fisher-Hartwig symbols, and related objects which have applications to Parity-time symmetric materials, Biophysical systems, and Quantum systems {\color{blue}[44]}.

\subsection{This paper's contributions}

\noindent In providing several comparisons between two, and three, dimensional vertex models, the forthcoming effort exhibits, irrespective of the total number of configurations in the sample spaces of the 7-vertex and 20-vertex models, that: generalizations of Solid-on-Solid models can be analyzed through applying a higher-dimensional analog of intertwining operations; the factorization of the universal R-matrix, previously characterized by the author in adaptations of the seminal Quantum Inverse Scattering framework, reflects upon hybrid integrability, and pure integrability, of the 20-vertex model; and finally, that admissible classes of boundary conditions can also reflect upon exactly solvable structure of Solid-on-Solid models that can be investigated in future work. Currently, from asymptotic properties of the transfer and quantum monodromy matrices of the 20-vertex model, the universal R-matrix plays a significant role in determining entries of the Boltzmann weight matrix of the Solid-on-Solid model, specifically from the fact that various exponentials of tensor products, which are later introduced as constitutents of the factorization for the universal R-matrix, are \textit{intertwined} under the action of the action of the Solid-on-Solid Boltzmann matrix. Establishing properties of such an intertwining action, in addition to the classes of admissible boundary conditions of each model, is of physical value for characterizing defects in materials, in addition to ferroelectric, and antiferroelectric, behaviors that two, and three, dimensional vertex models alike exhibit. To pursue directions of research interest as explained above, in the next section we introduce objects from each class of models.

\subsection{Motivation: integrability of inhomogeneous limit shapes, and of a Hamiltonian flow, of the 6-vertex model}

\noindent The 6-vertex model, as a two-dimensional vertex model, over the torus $\textbf{T}_{N} \equiv \big( V \big( \textbf{T}_N \big) , E \big( \textbf{T}_N \big) \big)$, the six-vertex model can be defined through the probability measure,

\begin{align*}
 \textbf{P}_{\textbf{T}_N} \big[ \omega \big] \equiv \textbf{P} \big[ \omega \big] \equiv     \frac{w \big( \omega \big)}{Z_{\textbf{T}_N}}           \text{, } 
\end{align*}

\noindent under domain-wall boundary conditions $\xi$,

\begin{align*}
 \textbf{P}^{\xi} \big[ \cdot \big] \equiv \textbf{P} \big[ \cdot \big]    \text{, }
\end{align*}

\noindent where $\omega$ is a \textit{six-vertex configuration} determined by the six possible configurations (see Figure 1 and Figure 2), with the weight in the numerator of the probability measure taking the form,

\begin{align*}
  w_{\mathrm{6V}} \big( \omega \big) \equiv w \big( \omega \big) \equiv a_1^{n_1} a_2^{n_2} b_1^{n_3} b_2^{n_4} c_1^{n_5} c_2^{n_6}  \text{, }
\end{align*}

\noindent for $a_1 , a_2 , b_1 , b_2 , c_1 , c_2 \geq 0$, with the partition function,

\begin{align*}
 Z_{\textbf{T}_N} \big( \omega , \Omega \big)  \equiv Z_{\textbf{T}_N} = \underset{\omega \in \Omega ( \textbf{T}_N ) }{\sum} w \big( \omega \big)   \text{. } 
\end{align*}

\noindent Besides $\textbf{P}_{\textbf{T}_N} \big[ \cdot \big]$, the disorder parameter of the six-vertex model is of the form,

\begin{align*}
    \Delta \equiv \frac{a_1 a_2 + a_3 a_4 - a_5 a_6}{2 \sqrt{a_1 a_2 a_3 a_4}}    \text{. } 
\end{align*}

\noindent The disorder parameter above, which is parameterized in weights $a_1, a_2, a_3, a_4, a_5, a_6$, gives rise to several behaviors of the 6-vertex model, ranging from disordered, and ordered, regimes, in addition to ferroelectric, and antiferroelectric, properties. For non-symmetric weights, the weights of the six-vertex model admit the parametrization,

\begin{align*}
   a_1 \equiv      a \text{ }  \mathrm{exp} \big(  H + V \big)   \text{, } \\ 
   a_2 \equiv   a  \text{ }  \mathrm{exp} \big( - H - V \big)   \text{, } \\  b_1 \equiv  \text{ }  \mathrm{exp} \big( H - V \big)    \text{, } \\ b_2 \equiv \text{ }  \mathrm{exp} \big( - H + V \big)   \text{, } \\ c_1 \equiv  c \lambda  \text{, } \\ c_2 \equiv c \lambda^{-1} \text{, } 
\end{align*}

\noindent for $a_1 \equiv a_2 \equiv a$, $b_1 \equiv b_2 \equiv b$, $c_1 \equiv c_2 \equiv c$, and $\lambda \geq 1$, and external fields $H,V$. From such a parametrization of the weights as given above, one can form the so-called $R$-matrix, for the standard basis of $\textbf{C}^2$, with,

\[
R \equiv R \big( u , H , V \big) \equiv 
  \begin{bmatrix}
      a \text{ }  \mathrm{exp} \big(  H + V \big)    & 0 & 0 & 0  \\
    0 & b \text{ } \mathrm{exp} \big( H - V \big) & c & 0  \\0 & c & b \text{ }  \mathrm{exp} \big( - H + V \big) & 0 \\ 0 & 0 & 0 & a \text{ }  \mathrm{exp} \big( - H - V \big) \\ 
  \end{bmatrix} \text{, }
\]

\noindent in the tensor product basis $e_1 \otimes e_1$, $e_1 \otimes e_2$, $e_2 \otimes e_1$, $e_2 \otimes e_2$, for $e_1 \equiv \big[ 1 \text{  } 0 \big]^{\mathrm{T}}$ and $e_1 \equiv \big[ 0 \text{  } 1 \big]^{\mathrm{T}}$. As a hallmark property of integrable systems, the R-matrix for the 6-vertex model above satisfies the Yang-Baxter equation,

\begin{align*}
R_{12} \big( u \big) R_{13} \big( u + v \big) R_{23} \big( v \big) =  R_{23} \big( v \big) R_{13} \big( u + v \big) R_{12} \big( u \big)    \text{. } 
\end{align*}

\noindent From classes of 6-vertex height functions, as well as another periodic function $\pi \big( x \big)$ in $x$, the pair $\big(  \pi  \big( x \big) , h\big( x \big) \big)$ can be identified with the cotangent space $T^{*} \mathcal{H}_{L,q}$, while the flow of the Hamiltonian can be identified with the pair $\big( \pi \big( x , y \big) , h \big( x , y \big) \big)$, in which,

\begin{align*}
        H_u \big( \pi \big( x , y \big) , h \big( x , y \big) \big) \equiv H_u \big( \pi , h \big) = {\int}_{[0,L]}   \mathcal{H}_{u - v ( x) } \big[  \partial_x h \big( x \big)  ,    \pi \big( x \big)   \big]  \mathrm{d} x    \text{, } 
\end{align*}

\noindent over $T^{*} \mathcal{H}_{L,q}$, for a solution $h \big( x , y \big)$ to the Euler Lagrange equations. Under the integral over $x$ provided above, $\mathcal{H}_u$ is the semigrand canonical free energy,

\begin{align*}
  \mathcal{H}_u \big( q , H \big)  \equiv  \mathcal{H}_u = \mathrm{log} \big[   Z_{\textbf{T}_{MN}}^n \big( u , H \big)       \big]               \text{, } 
\end{align*}

\noindent which can alternatively be expressed as the maximum over $\pm$, with,

\begin{align*}
  \mathcal{H}_u \big( q , H \big) \equiv \underset{\pm}{\mathrm{max}}    \text{ } \mathcal{H}^{\pm}_u \big( q , H \big) \equiv \underset{\pm}{\mathrm{max}}  \big\{     \pm H + l_{\pm} + \int_C  \psi^{\pm}_u \big( \alpha \big) \rho \big( \alpha \big)   \text{ } \mathrm{d} \alpha   \big\}     \text{, } \tag{$\mathrm{H}$}
\end{align*}

\noindent with $l_{-} \equiv \mathrm{log} \text{ } \mathrm{sinh} \big( \eta - u \big)$, $l_{+} \equiv \mathrm{log} \text{ } \mathrm{sinh} \text{ }  u$, and the density,

\begin{align*}
  \rho \big( \alpha \big) \equiv       \# \big\{    \text{Bethe roots along contours } C        \big\}   \equiv \underset{\alpha > 0 }{\bigcup}  \big\{ \alpha : \alpha \cap C \neq \emptyset \big\} \equiv  \bigg|  \big\{ \alpha : \alpha \cap C \neq \emptyset \big\}      \bigg| \text{. } 
\end{align*}

\noindent To connect the cotangent space $T^{*}_{\phi_1} \mathcal{H}_{q, l}$ with $T^{*}_{\phi_2} \mathcal{H}_{q, l}$ at time $T$ given the initial flow line $\big( \pi_0 , h_0 \big)$, one determines the unique critical point of the functional,

\begin{align*}
   S \big( \pi , h \big)  \equiv     S    =   {\int}_{[0,L]} \text{ }  {\int}_{[0,T]}    \big[                \pi \big( x , y \big) \partial_y h \big( x , y \big) - \mathcal{H}_{u - v ( x ) } \big( \partial_x h \big( x , y \big) , \pi \big( x , y \big) \big)             \big]    \mathrm{d} y \text{ }  \mathrm{d} x          \text{, } 
\end{align*}

\noindent which is $\big( \pi_0 , h_0 \big)$. Several computations with the Poisson bracket $\big\{ \cdot , \cdot \big\}$ are leveraged to characterize integrability of vertex models, which satisfies the following set of properties, given test functions $f$, $g$ and $h$:

\begin{itemize}
    \item [$\bullet$] \textit{Anticommutativity}. $\big\{ f, g \big\}  =  - \big\{ g , f \big\} $

    \item[$\bullet$] \textit{Bilinearity}. For real $a,b$, $\big\{ af + bg , h \big\} = a \big\{ f ,h \big\} + b \big\{ g , h \big\},$ and $\big\{ h , af + bg \big\} = a \big\{ h , f \big\} + b \big\{ h , g \big\} $

    \item[$\bullet$] \textit{Leibniz' rule}. $\big\{ fg , h \big\} = \big\{ f , h \big\} g + f \big\{ g , h \big\}$

    \item[$\bullet$] \textit{Jacobi identity}. $\big\{ f , \big\{ g , h \big\} \big\} + \big\{ g , \big\{ h , f \big\} \big\} + \big\{ h , \big\{ f , g \big\} \big\} = 0$ \end{itemize}

\noindent The action of the bracket above also admits the representation,

\begin{align*}
 \big\{ F ,  G \big\}   \equiv    i \int_{[-L,L]} \bigg[          \frac{\delta F}{\delta \psi} \frac{\delta G}{\delta \bar{\psi}} - \frac{\delta F}{\delta \bar{\psi}} \frac{\delta G}{\delta \psi }             \bigg]  \mathrm{d} x      \text{, } 
\end{align*}

\noindent for functionals $F$ and $G$, while the tensor product of the Poisson bracket takes the form,

\begin{align*}
 \big\{  A \overset{\bigotimes}{,} B \big\} \equiv     i \int_{[-L,L]} \bigg[  \frac{\delta A}{\delta \psi} \bigotimes \frac{\delta B}{\delta \bar{\psi}} - \frac{\delta A}{\delta \bar{\psi}} \bigotimes  \frac{\delta B}{\delta \psi }                     \bigg]    \mathrm{d} x     \text{, } 
\end{align*}

\noindent for test functionals $A$ and $B$. The transfer matrix, in addition to the quantum monodromy matrix, are central objects of study not only in Integrable Probability, but also in scattering methods. Before taking the weak infinite volume limit, the transfer matrix takes the form,

\begin{align*}
T_N \big( \lambda \big)  \equiv    E \big( -x , \lambda \big) T \big( x , y , \lambda \big) E \big( y , \lambda \big)   \text{, } 
\end{align*}

\noindent In weak infinite volume, the transfer matrix takes the form,

\begin{align*}
\underset{N \longrightarrow + \infty}{\mathrm{lim}} T_N \big( \lambda \big)  \equiv   T \big( \lambda \big)   = \underset{y \longrightarrow - \infty}{\underset{x \longrightarrow + \infty}{\mathrm{lim}} } E \big( -x , \lambda \big) T \big( x , y , \lambda \big) E \big( y , \lambda \big)   \text{, } 
\end{align*}

\noindent for the matrix exponential,

\begin{align*}
  E \big( x , \lambda \big) \equiv \mathrm{exp} \big( \lambda x U_1 
 \big) \text{, } 
\end{align*}

\noindent and also from the clockwise exponential action,

\begin{align*}
     T_L \big( \lambda \big) \equiv {\mathrm{exp}}  \bigg[ \underset{\mathrm{clockwise\text{ }  action}}{\int_{[-L,L]}} U \big( x , \lambda \big) \mathrm{d} x  \bigg]    \text{, } 
\end{align*}

\noindent for a real parameter $\lambda$, with the integrand $U$, a $2 \times 2$ matrix, and $N \geq L$, which satisfies the first order PDE,

\begin{align*}
   \frac{\partial F}{\partial x} =   U \big( x , \lambda \big)    F    \text{, } 
\end{align*}

\noindent for the vector valued function $F \big( x , t \big) \equiv F \equiv \big[ f_1 , f_2 \big]^{\mathrm{T}}$. For a solution $\psi$ to the Nonlinear Schrodinger's equation,

\begin{align*}
   i \frac{\partial \psi}{\partial t} = - \frac{\partial^2 \psi}{\partial x^2} + 2 \chi \big| \psi \big|^2 \psi      \text{, } 
\end{align*}

\noindent and real $\chi$, the nonconstant prefactor for $F$ takes the form of the linear combination,

\begin{align*}
  U = U_0 + \lambda U_1  \text{, } 
\end{align*}

\noindent

\noindent with,

\[
U_0  \equiv \sqrt{\chi}
  \begin{bmatrix}
       0  &  \psi \\
    \bar{\psi}  & 0 \text{ }  
  \end{bmatrix} \text{, } 
\]

\[
U_1  \equiv 
 \frac{1}{2i} \begin{bmatrix}
   1     &  0  \\
   0  &  -1 \text{ }  
  \end{bmatrix} \text{. } 
\]

\subsection{Discrete Probabilistic objects}

\subsubsection{20-vertex model}

For boundary conditions $\xi$, either those with sufficiently flat slope, ie, flat, or domain walls, such as those introduced over the pentagonal lattice, {\color{blue}[15]}, the 20-vertex probability measure supported over the triangular lattice, $\textbf{T}$, takes the form,

\begin{align*}
   \textbf{P}^{20V , \xi}_{\textbf{T}} \big[ \cdot \big] \equiv  \textbf{P}^{20V}_{\textbf{T}} \big[ \cdot \big]   \text{, }
\end{align*}

\noindent which is explicitly given by the ratio of the vertex weight function and the partition function,

\begin{align*}
\textbf{P}^{20V}_{\textbf{T}}[      \omega         ]   \equiv \textbf{P}^{20V}[   \omega     ]     =  \frac{w_{20V}(\omega)}{Z^{20V}_{\textbf{T}}} \equiv \frac{w(\omega)}{Z_{\textbf{T}}} \text{, }
\end{align*}

\noindent for some vertex configuration $\omega \in \Omega^{20V}$ - the 20-vertex sample space, and weights similar to those introduced in the previous section for the 6-vertex model, namely, {\color{blue}[15]},

\begin{align*}
     w_0 \equiv   a_1 a_2 a_3   \text{, } 
\\
    w_1 \equiv   b_1 a_2 b_3   \text{, } \\    w_2    \equiv   b_1 a_2 c_3    \text{, } \\
   w_3 \equiv       a_1 b_2 b_3 + c_1 c_2 c_3  \text{, } \\     w_4 \equiv    c_1 a_2 a_3        \text{, } \\               w_5 \equiv   b_1 c_2 a_3   \text{, } \\  w_6 \equiv  b_1 b_2 a_3    \text{, } 
\end{align*}

\noindent and the partition function,

\begin{align*}
 Z_{\textbf{T}} \equiv \underset{\omega \in \Omega^{20V}}{\sum}  w \big( \omega \big) \text{. }
\end{align*}

\noindent As was the case for the two-dimensional vertex model, the 6-vertex model, introduced in the previous subsection, one can also introduce finite volume approximations to the transfer matrix before taking the weak infinite volume limit, from spectral parameters $u,v,w$, with,

\[
T  \big(   \underline{M} , N ,  \underline{\lambda_{\alpha}} , u , v , w \big) \equiv    \overset{\underline{M}}{\underset{\underline{j}=0}{\prod}}  \text{ }  \overset{-N}{\underset{i=0}{\prod}}  \bigg[ \mathrm{exp} \big( \lambda_3 ( q^{-2} \xi^{s_i} ) \big)  \bigg[   \begin{smallmatrix}     q^{D_i}       &    q^{-2} a_i q^{-D_i-D_j} \xi^{s-s_i}        &   a_i a_{j} q^{-D_i - 3D_j} \xi^{s - s_i - s_j}  \\ a^{\dagger}_i q^{D_i} \xi^{s_i} 
             &      q^{-D_i + D_j} - q^{-2} q^{D_i -D_j} \xi^{s}     &     - a_j q^{D_i - 3D_j} \xi^{s-s_j}  \\ 0  &    a^{\dagger}_j q^{D_j} \xi^{s_j} &  q^{-D_j} \\    \end{smallmatrix} \bigg]    \bigg] \text{. } 
\]

\noindent As $M \longrightarrow + \infty$,  $N \longrightarrow - \infty$, the finite volume approximation above, in weak infinite volume, takes the form,

\begin{align*}
T \big(   \underline{\lambda}  \big) \equiv T\big( + \infty , - \infty ,   \underline{\lambda_{\alpha}} , \big\{ u_i \big\} , \big\{ v_j  \big\} , \big\{ w_k \big\} \big) =   \underset{\underline{M} \longrightarrow + \infty}{\mathrm{lim}} \text{ }  \underset{N \longrightarrow - \infty}{\mathrm{lim}}     T  \big(   M , N , \lambda_{\alpha} ,  v  ,  u    \big)    \text{, } 
\end{align*}

\noindent with,

\begin{align*}
 \textbf{T}^{3D} \big( \underline{\lambda} \big)   \equiv    \underset{ N \longrightarrow -  \infty}{\underset{\underline{M} \longrightarrow + \infty}{\mathrm{lim}}} \mathrm{tr} \bigg[     \overset{\underline{M}}{\underset{j=0}{\prod}}  \text{ }  \overset{-N}{\underset{k=0}{\prod}} \mathrm{exp} \big( \lambda_3 ( q^{-2} \xi^{s^j_k} ) \big)    \bigg[ \begin{smallmatrix}     q^{D^j_k}       &    q^{-2} a^j_k q^{-D^j_k -D^j_{k+1}} \xi^{s-s^k_j}        &  *_2\\ \big( a^j_k \big)^{\dagger} q^{D^j_k} \xi^{s^j_k} 
             &     *_1   &     - a^j_k q^{D^j_k - 3D^j_{k+1}} \xi^{s-s^j_k}  \\ 0  &    a^{\dagger}_j q^{D^j_k} \xi^{s^j_k} &  q^{-D^j_k} \\    \end{smallmatrix} \bigg]      \bigg] 
\text{, }
\end{align*}

\noindent for,

\begin{align*}
  *_1 \equiv  q^{-D^j_k + D^j_{k+1}} - q^{-2} q^{D^j_k -D^j_{k+1}} \xi^{s}    \text{, } \\   *_2 \equiv a^j_k a^j_{k+1} q^{-D^j_k - 3D^j_{k+1}} \xi^{s - s^j_k - s^j_{k+1}}   \text{. }
\end{align*}

\subsubsection{Solid-on-Solid model}

Despite the fact that previous adaptations of seminal work in {\color{blue}[17]} have studied objects relating to the Bethe ansatz, and several closely related objects, {\color{blue}[21},{\color{blue}22},{\color{blue}23},{\color{blue}24},{\color{blue}25},{\color{blue}26},{\color{blue}27},{\color{blue}28]}, adaptations of the quantum inverse scattering framework provided by the author in {\color{blue}[46]} are primarily reliant upon a higher-dimensional analog of computations with an L-operator of the 6-vertex model analyzed in {\color{blue}[42]}. As a byproduct of the quantum inverse scattering framework, it was conjectured in {\color{blue}[25]} that integrability of limit shapes for the 6-vertex model should imply integrability of a Hamiltonian flow under the presence of inhomogeneities, which was resolved by the author. For the rational 7-vertex model, given an R-matrix, the system of equations from which the intertwining vectors can be explicitly read off from takes the form,

\begin{align*}
 R \big( u - v \big) \bigg(   \psi \big( u \big)^a_b \otimes \psi \big( v \big)^b_c \bigg)  = \underset{b^{\prime}}{\sum}    \bigg(      \psi \big( v \big)^a_{b^{\prime}} \otimes \psi \big( u \big)^{b^{\prime}}_c \bigg)  W       \text{, }
\end{align*}

\noindent for two spectral parameters $u$ and $v$, intertwining vectors $\psi$, points $a,b,c,b^{\prime}$, and Boltzmann weight matrix $W$. In the equality of intertwining vectors above, the fact that the R-matrix for the rational 7-vertex model is dependent upon the difference of spectral parameters, $u-v$, rather than upon one, or two spectral parameters, $u$ and $v$ can be reflected through the combinatorial factors of the q-exponentials, in addition to indicates of the summation which depend upon $r$, a spectral parameter introduced along the rows, or columns, of some finite volume of the triangular lattice. With such observations, from factorization of the Universal R-matrix, it will be demonstrated that the system of relations takes the form,

\[
(*) \equiv \left\{\!\begin{array}{ll@{}>{{}}l} (1):  \mathcal{R}_1 \beta_l \big( u \big) \beta_{l+1} \big( u \big) =  \beta_l \big( v \big) \beta_{l+1} \big( u \big)   W_1 
\text{, } \\   (2):   \mathcal{R}_1 \beta_l \big( u \big) \gamma_{l+1} \big( u \big) =    \beta_l \big( v \big) \gamma_{l+1} \big( u \big)    W_1       \text{, } \\      (3):  \mathcal{R}_1 \beta_l \big( u \big) Z_{l+1} \big( u \big) =   \beta_l \big( v \big) Z_{l+1} \big( u \big)    W_1   \text{, } \\    (4):  \mathcal{R}_1 \beta_l \big( u \big) \beta_{l+1} \big( u \big) =   \gamma_l \big( v \big) \beta_{l+1} \big( u \big) W_1  \text{, } \\   (5):           \mathcal{R}_1 \gamma_l \big( u \big) \gamma_{l+1} \big( v \big) =  \gamma_l \big( v \big) \gamma_{l+1} \big) u \big)  W_1  \text{, } \\   (6): \mathcal{R}_1 \gamma_l \big( u \big) Z_{l+1} \big( v \big) =  \gamma_l \big( v \big) Z_{l+1} \big( u \big) W_1  \text{, } \\ (7): \mathcal{R}_1  Z_l \big( u \big) \beta_{l+1} \big( v \big) =  Z_l \big( v \big) \beta_{l+1} \big( u \big) W_1  \text{, } \\ (8): \mathcal{R}_1 Z_l \big( u \big) \gamma_{l+1} \big( v \big) =  Z_l \big( v \big) \gamma_{l+1} \big( u \big) W_1 \text{, } \\ (9): \mathcal{R}_1 Z_l \big( u \big) Z_{l+1} \big( v \big) =  Z_l \big( v \big) Z_{l+1} \big( u \big)      W_1 \text{, }       \end{array}\right.
\]

\noindent for the 20-vertex intertwining vectors,

\begin{align*}
  X_l \big( u \big) = \bigg[ \begin{smallmatrix}   \beta_l ( u )  \\ \gamma_l ( u ) \\ Z_l ( u ) 
  \end{smallmatrix} \bigg]  \text{, } \\ X_{l+1} \big( u \big) = \bigg[ \begin{smallmatrix}   \beta_{l+1} ( u )  \\ \gamma_{l+1} ( u ) \\ Z_{l+1} ( u ) 
  \end{smallmatrix} \bigg]  \text{. } 
\end{align*}

\noindent To distribute terms from the factorization of the Universal R-matrix, with $X_l \big( u \big)$ and $X_{l+1} \big( u \big)$ above, observe that from the first term of the R-matrix from the rational 7-vertex model,

\begin{align*}
 R \big( u - v \big)    \text{, }
\end{align*}

\noindent appears as a prefactor to the first intertwining vector $\psi \big( u \big)^a_b$, which appears in the tensor product,

\begin{align*}
  \psi \big( u \big)^a_b \otimes \psi \big( v \big)^b_c     \text{, }
\end{align*}

\noindent of intertwining vectors. For intertwining vectors of the 20-vertex model, in comparison to those of the rational 7-vertex model, the q-exponential factor,

\begin{align*}
 \underset{m \in \textbf{N}
}{\underset{\gamma \in \Delta+ ( A)}{\prod}}    \mathrm{exp}_q \big[ \big( q - q^{-1} \big)          s^{-1}_{m , \delta - \gamma} e_{\delta - \gamma + m \delta} \otimes f_{ \delta - \gamma + m \delta}              \big]  \text{, }
\end{align*}

\noindent appearing in the the Universal R-matrix factorization is distributed to $\psi \big( u \big)^a_b$ before taking the resultant tensor product with the remaining intertwining vector $\psi \big( v \big)^b_c$. From previous work on the rational 7-vertex model that has been discussed, each entry of the intertwining vectors, over the square lattice, has a $1$ either in the first or second entry, in addition to the other factor being dependent upon a product of strictly positive parameters, with contributions from $\alpha$, $n$, and $l$. From the system of equations for the rational 7-vertex model, the explicit form of the three entries appearing in the intertwining vectors of the 20-vertex model takes a similar form. In obtaining a general solution from the system of nine relations above, one distributes the tensor product from the intertwining vectors,

\begin{align*}
     X_l \big( u \big)    \text{, } \\ X_{l+1} \big( u \big) 
\text{. }
\end{align*}

\noindent In the system for the 20-vertex SOS model that is dependent upon entries of the Boltzmann weight matrix, the order in which spectral parameters are introduced into the tensor product, from each intertwining vector, is reversed. That is, from each of the nine relations listed above for each entry of the R-matrix, and of the Boltzmann weight matrix, the sequence in which the intertwining vectors, which are respectively dependent upon $u \equiv \underline{u}$ and $v\equiv \underline{v}$, comprise the reversed transformation that is applied to the Boltzmann weight matrix. Altogether, the universal R-matrix factorization into q-exponential, the K-matrix, and spectral parameters, implies that one must consider exponentials of the form,

\begin{align*}
   \mathscr{A}_1 \mathscr{A}_2 \mathscr{A}_3 \mathscr{A}_4 \text{, }
\end{align*}

\noindent where,

\begin{align*}
      \mathscr{A}_1 \equiv  \mathrm{exp} \bigg[         \hbar   \big( q - q^{-1} \big) s^{-1}_{m,\gamma} e_{\gamma+m \delta} \otimes f_{\gamma+ m \delta} \bigg] \text{, } \\  \mathscr{A}_2 \equiv  \mathrm{exp} \bigg[             \big( q - q^{-1} \big) \underset{m \in \textbf{Z}^+}{\sum}            \overset{r}{\underset{i^{\prime} \neq j^{\prime} \in \textbf{Z}}{\underset{i-j,i^{\prime}-j^{\prime}=1}{\sum}}}          u_m \big( i - j\big) \big( i^{\prime} - j^{\prime} \big)     e_{m\delta,\alpha_{i-j}} \otimes f_{m\delta , a_{j-j^{\prime}}}                       \bigg] \text{, } \\  \mathscr{A}_3 \equiv          \underset{m \in \textbf{N}}{\underset{\gamma \in \Delta_+ ( A)}{\prod}}         \mathrm{exp} \bigg[ \big( q - q^{-1} \big)        s^{-1}_{m,\delta-\gamma} e_{\delta - \gamma + m \delta }       \otimes   f_{\delta-\gamma+m \delta}    \bigg]  \text{, } \\ \mathscr{A}_4 \equiv          \mathrm{exp} \bigg[ \hbar    \overset{r}{\underset{i^{\prime}-j^{\prime}\in \textbf{Z}}{\underset{i-j \in \textbf{Z}}{\underset{i-j, i^{\prime}-j^{\prime}=1}{\sum} }}}     \beta_{(i-j)(i^{\prime}-j^{\prime})} h_{\alpha ( i-j)}  \otimes h_{\alpha(i^{\prime}-j^{\prime})}   \bigg]                     \text{. }
\end{align*}

\noindent In comparison to the first expression introduced for the exponential of the K-matrix that is dependent upon a \textit{single} spectral parameter, rather than the \textit{difference} of two spectral parameters, exponentials of the form above are introduced for boundary conditions to the first term $\mathcal{R} \big( u - v \big)$ appearing in the 20-vertex intertwining relation. Following the overview in the next subsection, we demonstrate how a system of relations, from the nine components, are obtained from the intertwining vectors.

\subsection{Paper overview}

Beginning in the next section, we demonstrate how the intertwining relation for the rational 7-vertex model,

\begin{align*}
 R \big( u - v \big) \bigg(  \psi \big( u \big)^a_b \otimes \psi \big( v \big)^b_c \bigg)  = \underset{b^{\prime}}{\sum}      \bigg(    \psi \big( v \big)^a_{b^{\prime}} \otimes \psi \big( u \big)^{b^{\prime}}_c \bigg)  W       \text{, }
\end{align*}

\noindent can be applied to the 20-vertex model, which as mentioned in the previous subsection, takes the form,

\begin{align*}
 \mathcal{R} \big( u - v \big) \bigg(   \psi^{20V} \big( u \big)^a_b \otimes \psi^{20V} \big( v \big)^b_c \bigg) = \underset{b^{\prime}}{\sum}     \bigg(     \psi^{20V} \big( v \big)^a_{b^{\prime}} \otimes \psi^{20V} \big( u \big)^{b^{\prime}}_c \bigg)  W^{20V}       \text{, }
\end{align*}

\noindent Moving forwards, we denote the SOS 20-vertex Boltzman matrix with $W^{20V} \equiv W$. After having obtained each equation from the system, we derive the form of the intertwining vectors for the 20-vertex model. The entries of the two intertwining vectors for the SOS 20-vertex model, as in the case for the two intertwining vectors of the rational SOS 7-vertex model, depend upon multiplying equations from the system of nine relations together. When taking the product of different equations within the system, we make use of the following sequence identifications between intertwining vectors,

\[
\left\{\!\begin{array}{ll@{}>{{}}l}    \beta_l \big( v \big) \beta_{l+1} \big( u \big) \longleftrightarrow \beta_{l+1} \big( v \big) \beta_l \big( u \big)   \text{, }   \\  \beta_l \big( v \big) \gamma_{l+1} \big( u \big) \longleftrightarrow \gamma_{l+1} \big( v \big) \beta_l \big( u \big)   \text{, }              \\     \beta_l \big( v \big) Z_{l+1} \big( u \big)     \longleftrightarrow  Z_{l+1} \big( v \big) \beta_l \big( u \big)  \text{, }         \\     \gamma_l \big( v \big) \beta_{l+1} \big( u \big)                 \longleftrightarrow         \beta_{l+1} \big) v \big) \gamma_l \big( u \big)           \text{, }         \\  \gamma_l \big( v \big) \gamma_{l+1} \big( u \big) \longleftrightarrow  \gamma_{l+1} \big( v \big) \gamma_l \big( u \big)  \text{, }         \\    \gamma_l \big( v \big) Z_{l+1} \big( u \big)   \longleftrightarrow Z_{l+1} \big( v \big)  \gamma_l \big( u \big)   \text{, }         
\\ Z_l \big(  v \big) \beta_{l+1} \big( u \big)  \longleftrightarrow                \beta_{l+1} \big( v \big) Z_l \big( u \big)     \text{, }  \\   Z_l \big(  v \big)             \gamma_{l+1} \big( u \big) 
 \longleftrightarrow                \gamma_{l+1} \big( v \big) Z_l \big( u \big)           \text{, }  \\  Z_l \big(  v \big)            Z_{l+1} \big( u \big)   \longleftrightarrow    Z_{l+1} \big( v \big) Z_l \big( u \big)         \text{. }  
\end{array}\right.
\] 

\noindent In the series of associations above between intertwining vectors, the order in which spectral parameters appear is maintained, while the order in which the entry of the intertwining vector appears is reversed. In comparison to the lower dimensional system of equations appearing for the SOS rational 7-vertex model that is manipulated by Antonenko and Valinevish, {\color{blue}[1]}, the factorization of the Universal R-matrix implies that one must consider the following set of linear combinations,

\begin{align*}
\underset{k \in \textbf{N}: k < r}{\mathrm{span}} \bigg\{  \mathscr{R}_1 \big( k \big) \mathscr{R}_2 \big( k \big) \mathscr{R}_3 \big( k \big) \mathscr{R}_4 \big( k \big)  \bigg\}   \equiv   \underset{k \in \textbf{N}: k < r}{\mathrm{span}} \bigg\{  \mathscr{R}_1  \mathscr{R}_2  \mathscr{R}_3 \mathscr{R}_4 \big( k \big)  \bigg\}      \text{, }
\end{align*}

\noindent where,

\begin{align*}
      \mathscr{R}_1 \equiv  \underset{m \in \textbf{N}}{\underset{\gamma \in \Delta_+ ( A)}{\prod}} \mathrm{exp}_q \bigg[ ( q - q^{-1} ) s^{-1}_{m,\gamma} e_{\gamma+m\delta} \otimes f_{\gamma + m \delta}  \bigg]   \text{, } \\ \mathscr{R}_2 \equiv    \mathrm{exp} \bigg[ ( q - q^{-1} ) \underset{m \in \textbf{Z}}{\sum} \text{ }  \overset{r}{\underset{i-j, i^{\prime}-j^{\prime}=1}{\sum}}                u_m ( i - j ) ( i^{\prime} - j^{\prime} ) e_{m\delta , \alpha_{i-j}}  \otimes f_{m\delta , \alpha_{i^{\prime}-j^{\prime}}}                    \bigg]  \text{, } \\ \mathscr{R}_3 \equiv      \underset{ m \in \textbf{N}}{\underset{\gamma \in \Delta_+ ( A)}{\prod}}     \mathrm{exp}_q \bigg[ \big( q - q^{-1} \big) s^{-1}_{m,\gamma-\delta} e_{\delta-\gamma+m \delta} \otimes f_{\delta-\gamma + m \delta}   \bigg]  \text{, } \\ \mathscr{R}_4 \equiv      \underset{ m \in \textbf{N}}{\underset{\gamma \in \Delta_+ ( A)}{\prod}}   \mathrm{exp} \bigg[ \hbar              \overset{r}{\underset{j\neq j^{\prime}}{\underset{i-j,i-j^{\prime}=1}{\sum}}}   \beta_{(i-j)(i-j^{\prime})} h_{\alpha(i-j)}   \otimes h_{\alpha(i-j^{\prime})}   \bigg]         \text{. }
\end{align*}

\noindent Entrywise, the vector above plays the role of the Universal R-matrix,

\begin{align*}
\mathcal{R} \big( u - v \big)     \text{, }
\end{align*}

\noindent from the following equality between R-matrices, and Boltzmann weight matrices,

\begin{align*}
    \mathcal{R} \big( u - v \big) \bigg(   \psi^{20V} \big( v \big)^a_{b^{\prime}} \otimes \psi^{20V} \big( u \big)^{b^{\prime}}_c     \bigg)    \equiv \bigg(  \mathcal{R} \big( u - v \big)  \psi^{20V} \big( v \big)^a_{b^{\prime}}   \bigg)  \otimes \psi^{20V} \big( u \big)^{b^{\prime}}_c     \text{. }
\end{align*}

\noindent Deducing that the LHS of equations in the system, as will be shown in the next section, for each component, reduces to,

\begin{align*}
 \bigg[ \begin{smallmatrix}   \beta_l ( v )  \\ \gamma_l ( v ) \\ Z_l ( v ) 
  \end{smallmatrix} \bigg]   \otimes  \bigg[ \begin{smallmatrix}   \beta_l ( u )  \\ \gamma_l ( u ) \\ Z_l ( u ) 
  \end{smallmatrix} \bigg]   \Longleftrightarrow   \mathrm{span} \bigg\{ v : \beta_l \big( v \big) \otimes \bigg[ \begin{smallmatrix}    \beta_l ( u )  \\ \gamma_l ( u ) \\ Z_l ( u )  
  \end{smallmatrix} \bigg]  \bigg\}   \bigcup    \mathrm{span} \bigg\{ v : \gamma \big( v \big) \otimes \bigg[ \begin{smallmatrix}    \beta_l ( u )  \\ \gamma_l ( u ) \\ Z_l ( u )  
  \end{smallmatrix} \bigg]  \bigg\}    \bigcup  \mathrm{span} \bigg\{  v : Z_l \big( v \big) \\   \otimes \bigg[ \begin{smallmatrix}    \beta_l ( u )  \\ \gamma_l ( u ) \\ Z_l ( u )  
  \end{smallmatrix} \bigg]  \bigg\}      \text{. }
\end{align*}

\section{intertwining vectors, and the Boltzmann weight matrix, from the SOS 20-vertex model}

\noindent Before proving the main item of interest, we discuss characteristics of the system of equations. From the previous nine relations in (*), the generalized system across all entries of the Universal R-matrix takes the form,

\[
\left\{\!\begin{array}{ll@{}>{{}}l} (1):  \mathcal{R} \beta_l \big( u \big) \beta_{l+1} \big( u \big) =  \beta_l \big( v \big) \beta_{l+1} \big( u \big)  W  
\text{, } \\   (2):   \mathcal{R} \beta_l \big( u \big) \gamma_{l+1} \big( u \big) =    \beta_l \big( v \big) \gamma_{l+1} \big( u \big)    W       \text{, } \\      (3):  \mathcal{R} \beta_l \big( u \big) Z_{l+1} \big( u \big) =   \beta_l \big( v \big) Z_{l+1} \big( u \big)   W    \text{, } \\    (4):  \mathcal{R} \beta_l \big( u \big) \beta_{l+1} \big( u \big) =  \gamma_l \big( v \big) \beta_{l+1} \big( u \big) W   \text{, } \\   (5):           \mathcal{R} \gamma_l \big( u \big) \gamma_{l+1} \big( v \big) =  \gamma_l \big( v \big) \gamma_{l+1} \big) u \big)   W \text{, } \\   (6): \mathcal{R} \gamma_l \big( u \big) Z_{l+1} \big( v \big) =  \gamma_l \big( v \big) Z_{l+1} \big( u \big)  W \text{, } \\ (7): \mathcal{R}  Z_l \big( u \big) \beta_{l+1} \big( v \big) =  Z_l \big( v \big) \beta_{l+1} \big( u \big) 
 W \text{, } \\ (8): \mathcal{R} Z_l \big( u \big) \gamma_{l+1} \big( v \big) =  Z_l \big( v \big) \gamma_{l+1} \big( u \big) W  \text{, } \\ (9): \mathcal{R} Z_l \big( u \big) Z_{l+1} \big( v \big) =  Z_l \big( v \big) Z_{l+1} \big( u \big)   W \text{. }          \end{array}\right.
\] 

\noindent The matrix valued system of nine relations above contains one hundred and thirfty five more relations than (*). Furthermore, the system of relations above, as a higher-dimensional analog of the system of relations obtained from the R-matrix of the rational 7-vertex model in {\color{blue}[1]}, decomposes as follows. The first relation from the system reads,

\begin{align*}
    \mathcal{R} \beta_l \big( u \big) \beta_{l+1} \big( u \big) = W \beta_l \big( v \big) \beta_{l+1} \big( u \big)   \Longleftrightarrow  \begin{bmatrix} \mathcal{R}_{(1,1)} & \mathcal{R}_{(2,1)} & \mathcal{R}_{(3,1)} & \mathcal{R}_{(4,1)} \\ \mathcal{R}_{(1,2)} & \mathcal{R}_{(2,2)} & \mathcal{R}_{(3,2)} & \mathcal{R}_{(4,2)} \\ \mathcal{R}_{(1,3)} & \mathcal{R}_{(2,3)} & \mathcal{R}_{(3,3)} & \mathcal{R}_{(4,1)} \\ \mathcal{R}_{(1,4)} & \mathcal{R}_{(2,4)} & \mathcal{R}_{(3,4)} & \mathcal{R}_{(4,4)}
    \end{bmatrix} \\ \times \beta_l \big( u \big) \beta_{l+1} \big( u \big)   =  \beta_l \big( v \big)  \beta_{l+1} \big( u \big)     \begin{bmatrix} W_{(1,1)} & W_{(2,1)} & W_{(3,1)}  & W_{(4,1)} \\ W_{(1,2)} & W_{(2,2)} & W_{(3,2)} & W_{(4,2)} \\ W_{(1,3)} & W_{(2,3)} & W_{(3,3)} & W_{(4,3)} \\ W_{(1,4)} & W_{(2,4)} & W_{(3,4)} & W_{(4,4)}
    \end{bmatrix}  
\text{, } 
\end{align*}    

\noindent which itself decomposes into sixteen relations,

\[
(*)(1) \equiv \left\{\!\begin{array}{ll@{}>{{}}l}
(1): \mathcal{R}_{(1,1) }  \beta_l \big( u \big) \beta_{l+1} \big( v \big) =  \beta_l \big( v \big) \beta_{l+1} \big( u \big) W_{(1,1)}  \text{, } \\ (2): \mathcal{R}_{(1,2) }  \beta_l \big( u \big) \beta_{l+1} \big( v \big) = \beta_l \big( v \big) \beta_{l+1} \big( u \big)  W_{(1,2)}  \text{, }  \\ (3):        \mathcal{R}_{(1,3) }  \beta_l \big( u \big) \beta_{l+1} \big( v \big) = \beta_l \big( v \big) \beta_{l+1} \big( u \big)   W_{(1,3)}  \text{, }      \\ (4):    \mathcal{R}_{(1,4) }  \beta_l \big( u \big) \beta_{l+1} \big( v \big) =  \beta_l \big( v \big) \beta_{l+1} \big( u \big) W_{(1,4)}  \text{, }  \\ (5):     \mathcal{R}_{(2,1) }  \beta_l \big( u \big) \beta_{l+1} \big( v \big) =  \beta_l \big( v \big) \beta_{l+1} \big( u \big) W_{(2,1)}  \text{, } \\ (6):   \mathcal{R}_{(2,2) }  \beta_l \big( u \big) \beta_{l+1} \big( v \big) =  \beta_l \big( v \big) \beta_{l+1} \big( u \big) W_{(2,2)}  \text{, }  \\ (7):   \mathcal{R}_{(2,3) }  \beta_l \big( u \big) \beta_{l+1} \big( v \big) =  \beta_l \big( v \big) \beta_{l+1} \big( u \big) W_{(2,3)}  \text{, } \\ (8):   \mathcal{R}_{(2,4) }  \beta_l \big( u \big) \beta_{l+1} \big( v \big) = \beta_l \big( v \big) \beta_{l+1} \big( u \big) W_{(2,4)}   \text{, }  \\ (9):   \mathcal{R}_{(3,1) }  \beta_l \big( u \big) \beta_{l+1} \big( v \big) =  \beta_l \big( v \big) \beta_{l+1} \big( u \big) W_{(3,1)}  \text{, } \\ (10):   \mathcal{R}_{(3,2) }  \beta_l \big( u \big) \beta_{l+1} \big( v \big) =  \beta_l \big( v \big) \beta_{l+1} \big( u \big) W_{(3,2)}  \text{, } \\ (11):   \mathcal{R}_{(3,3) }  \beta_l \big( u \big) \beta_{l+1} \big( v \big) = \beta_l \big( v \big) \beta_{l+1} \big( u \big)   W_{(3,3)} \text{, } \\ (12):   \mathcal{R}_{(3,4) }  \beta_l \big( u \big) \beta_{l+1} \big( v \big) =  \beta_l \big( v \big) \beta_{l+1} \big( u \big) W_{(3,4)}  \text{, }  \\ (13):   \mathcal{R}_{(4,1) }  \beta_l \big( u \big) \beta_{l+1} \big( v \big) = \beta_l \big( v \big) \beta_{l+1} \big( u \big)  W_{(4,1)}   \text{, } \\ (14):   \mathcal{R}_{(4,2) }  \beta_l \big( u \big) \beta_{l+1} \big( v \big) = \beta_l \big( v \big) \beta_{l+1} \big( u \big)  W_{(4,2)}   \text{, }\\ (15):   \mathcal{R}_{(4,3) }  \beta_l \big( u \big) \beta_{l+1} \big( v \big) = \beta_l \big( v \big) \beta_{l+1} \big( u \big)  W_{(4,3)}  \text{, } \\  (16    ):   \mathcal{R}_{(4,4) }  \beta_l \big( u \big) \beta_{l+1} \big( v \big) =  \beta_l \big( v \big) \beta_{l+1} \big( u \big) W_{(4,4)}  \text{, }
\end{array}\right.
\]

\noindent where,

\begin{align*}
  \mathcal{R}_{(1,1)} \equiv   \mathcal{R}_{(1,1)} \big( u - v \big)  \Longleftrightarrow W_{(1,1)} \equiv W_{(1,1)} \big( u - v \big)    \text{, } \\  \mathcal{R}_{(1,2)} \equiv   \mathcal{R}_{(1,2)} \big( u - v \big) \Longleftrightarrow W_{(1,2)} \equiv W_{(1,2)} \big( u - v \big)    \text{, } \\ \mathcal{R}_{(1,3)} \equiv   \mathcal{R}_{(1,3)} \big( u - v \big)  \Longleftrightarrow W_{(1,3)} \equiv W_{(1,3)} \big( u - v \big)   \text{, } \\ \mathcal{R}_{(1,4)} \equiv   \mathcal{R}_{(1,4)} \big( u - v \big)  \Longleftrightarrow W_{(1,4)} \equiv W_{(1,4)} \big( u - v \big)   \text{, } \\ \mathcal{R}_{(2,1)} \equiv   \mathcal{R}_{(1,2)} \big( u - v \big) \Longleftrightarrow W_{(2,1)} \equiv W_{(2,1)} \big( u - v \big)   \text{, }  \\ \mathcal{R}_{(2,2)} \equiv   \mathcal{R}_{(2,2)} \big( u - v \big)  \Longleftrightarrow W_{(2,2)} \equiv W_{(2,2)} \big( u - v \big)   \text{, }  \\ \mathcal{R}_{(2,3)} \equiv   \mathcal{R}_{(2,3)} \big( u - v \big) \Longleftrightarrow W_{(2,3)} \equiv W_{(2,3)} \big( u - v \big)    \text{, } \\ \mathcal{R}_{(2,4)} \equiv   \mathcal{R}_{(2,4)} \big( u - v \big) \Longleftrightarrow W_{(2,4)} \equiv W_{(2,4)} \big( u - v \big)   \text{, }   \\ \mathcal{R}_{(3,1)} \equiv   \mathcal{R}_{(3,1)} \big( u - v \big)  \Longleftrightarrow W_{(3,1)} \equiv W_{(3,1)} \big( u - v \big)   \text{, }  \\ \mathcal{R}_{(3,2)} \equiv   \mathcal{R}_{(3,2)} \big( u - v \big)  \Longleftrightarrow W_{(3,2)} \equiv W_{(3,2)} \big( u - v \big)   \text{, } \\ \mathcal{R}_{(3,3)} \equiv   \mathcal{R}_{(3,3)} \big( u - v \big) \Longleftrightarrow W_{(3,3)} \equiv W_{(3,3)} \big( u - v \big)   \text{, }  \\ \mathcal{R}_{(3,4)} \equiv   \mathcal{R}_{(3,4)} \big( u - v \big) \Longleftrightarrow W_{(3,4)} \equiv W_{(3,4)} \big( u - v \big)   \text{, }  \\ \mathcal{R}_{(4,1)} \equiv   \mathcal{R}_{(4,1)} \big( u - v \big) \Longleftrightarrow W_{(4,1)} \equiv W_{(4,1)} \big( u - v \big)   \text{, } \\ \mathcal{R}_{(4,2)} \equiv   \mathcal{R}_{(4,2)} \big( u - v \big) \Longleftrightarrow W_{(4,2)} \equiv W_{(4,2)} \big( u - v \big)   \text{, }  \\ \mathcal{R}_{(4,3)} \equiv   \mathcal{R}_{(4,3)} \big( u - v \big) \Longleftrightarrow W_{(4,3)} \equiv W_{(4,3)} \big( u - v \big)   \text{, }  \\ \mathcal{R}_{(4,4)} \equiv   \mathcal{R}_{(4,4)} \big( u - v \big) \Longleftrightarrow W_{(4,4)} \equiv W_{(4,4)} \big( u - v \big)    \text{. } 
\end{align*}

\noindent Each of the remaining eight relations from the extension of (*) decomposes into sixteen relations in the same way that (1) decomposes into the collection of sixteen relations above, into $(*)(2),\cdots,(*)(16)$. To determine the solution set,

\begin{align*}
  \mathcal{S}\mathcal{S} \equiv \underset{1 \leq i \leq 16}{\bigcup} (*)(i)  \text{, }
\end{align*}

\noindent of the union over systems (*)$(i)$, it suffices to argue that the general solution, independent of the entry of the Universal R-matrix, and Boltzmann weight, matrices, takes the form indicated below. Below, we provide the statement of the main result. In the argument we exhibit how the first system of relations in (*) is formed, from which the general solution to all equations of the system can be obtained.

\bigskip

\noindent In the items collected in the main result below, as a matter of notation, denote matrices of the form,

\begin{align*}
\bigg[ \begin{smallmatrix}
 *_1 & *^{\prime}_1 \\ *_2 & *^{\prime}_2 \\ *_3 & *^{\prime}_3 
\end{smallmatrix} \bigg| \underline{u} \bigg] \text{, }
\end{align*}

\noindent as the positions, $*_1$, $*_2$, and $*_3$, over the triangular lattice which the Boltzmann weight matrix $W$ acts on, which are dependent upon the spectral parameter $\underline{u}$, which are mapped onto $*^{\prime}_1$, $*^{\prime}_2$, and $*^{\prime}_3$, respectively. The mapping, 

\begin{align*}
  *_1 \mapsto *^{\prime}_1   \text{, } \\  *_2 \mapsto *^{\prime}_2 \text{, } \\  *_3 \mapsto *^{\prime}_3 \text{, } 
\end{align*}

\noindent performs a translation oof three points over the triangular lattice to another set of three points over the triangular lattice. In comparison to representations for the Boltzmann weight matrix for the rational 7-vertex model, those for the 20-vertex model are dependent upon the solution set $\mathcal{S} \mathcal{S}$.

\bigskip

\noindent \textbf{Theorem} (\textit{solution sets to the system of equations for the Boltzmann weight matrix of the SOS 20-vertex model}). There exists constants $\mathcal{C}_1, \mathcal{C}_2,\mathcal{C}_3$, and $\mathscr{C}_1$, for which the Boltzmann weight matrix, $W$, obtained from the intertwining relation equals,

\begin{align*}
  W \bigg[   \begin{smallmatrix}
     l + 2  &  l+1   \\ l + 1  & l + 1  \\  l & l + 1
  \end{smallmatrix}\bigg| \underline{u} \bigg] \equiv   W \bigg[   \begin{smallmatrix}
     l + 1  &  l   \\ l   & l   \\   l - 1  & l 
  \end{smallmatrix}\bigg| \underline{u} \bigg] =   \mathcal{C}_1 \equiv \mathcal{C}_1 \big( l , \underline{u} \big)  \equiv \mathcal{C}_1 \big( 1 , \big( i , j , r \big) \big)  
  \text{, } \\  W \bigg[   \begin{smallmatrix}
       l - 1   &  l  \\  l  & l \\ l + 1 &  l 
  \end{smallmatrix}\bigg| \underline{u} \bigg] \equiv  W \bigg[   \begin{smallmatrix}
       l    &  l - 1   \\  l - 1   & l - 1  \\ l  &  l - 1 
  \end{smallmatrix}\bigg| \underline{u} \bigg] =  \mathcal{C}_2 \equiv \mathcal{C}_2 \big( l , \underline{u} \big)  \equiv \mathcal{C}_2 \big( 1 , \big( i , j , r \big) \big)    \text{, } \\ W \bigg[   \begin{smallmatrix}
      l - 2   &  l + 1    \\  l - 1   &       l - 1    \\     l  &  l + 1     
  \end{smallmatrix}\bigg| \underline{u} \bigg] \equiv W \bigg[   \begin{smallmatrix}
      l - 1   &  l    \\  l    &       l     \\     l  + 1  &  l       
  \end{smallmatrix}\bigg| \underline{u} \bigg]  =  \mathcal{C}_3 \equiv \mathcal{C}_3 \big( l , \underline{u} \big)  \equiv \mathcal{C}_3 \big( 1 , \big( i , j , r \big) \big) 
 \text{, }
\end{align*}

\noindent from which the intertwining vectors take the form, given $a < b^{\prime} < c \in V \big( \textbf{T} \big)$,

\begin{align*}
 \psi^{20V} \big( u \big)^a_{b^{\prime}} \equiv       \bigg[     \begin{smallmatrix}
 1 \\ 1 \\   \mathscr{C}_1 ( l , \underline{u} )         \end{smallmatrix} \bigg]  \equiv  \bigg[     \begin{smallmatrix}
 1 \\ 1 \\   \mathscr{C}_1         \end{smallmatrix} \bigg]  
 \text{, } \\  \psi^{20V} \big( u \big)^{b^{\prime}}_c \equiv   \bigg[     \begin{smallmatrix}
 1 \\   \frac{1}{\mathscr{C}_1 ( l , \underline{u} )}  \\ 1   \end{smallmatrix} \bigg] \equiv   \bigg[     \begin{smallmatrix}
 1 \\   \frac{1}{\mathscr{C}_1 }   \\ 1 \end{smallmatrix} \bigg]    \text{. }
\end{align*}

\noindent \textit{Proof of Theorem}. For obtaining the solution to the SOS 20-vertex system of equations, we begin by listing out one side of the relations obtained in (*), which can then be used to obtain $\mathcal{S}\mathcal{S}$ for all entries of the Universal R-matrix. In particular, by resolving the tensor product,

\[
\mathcal{R} \bigg[ \begin{bmatrix}
     \beta_l \big( u \big) \\ \gamma_l \big( u \big) \\ Z_l \big( u \big)    \end{bmatrix} \otimes \begin{bmatrix}  \beta_{l+1} \big( v \big) \\ \gamma_{l+1} \big( v \big) \\ Z_{l+1} \big( v \big) \end{bmatrix} \bigg]  \text{, }
\]

\noindent into components one obtains the LHS of each equation in (*). For the remaining terms on the RHS, by resolving the tensor product,

\begin{align*}
     \bigg[  \begin{bmatrix} \beta_l \big(  v \big) \\ \gamma_l \big( v \big) \\ Z_l \big( v \big) \end{bmatrix} \otimes \begin{bmatrix} \beta_{l+1} \big( u \big) \\ \gamma_{l+1} \big( u \big) \\ Z_{l+1} \big( u \big) \end{bmatrix}  \bigg] W         \text{, }
\end{align*}

\noindent into components one obtains the RHS of each equation in (*). The collection of relations from (*) can be used to obtain the intertwining vectors, in addition to the Boltzmann weight matrix of the SOS 20-vertex model, from the observation that the prefactor to the entries of the system generated by the tensor product above takes the form,

\begin{align*}
 \mathscr{R}_1 \mathscr{R}_2 \mathscr{R}_3 \mathscr{R}_4    \text{, }
\end{align*}

\noindent where,

\begin{align*}
      \mathscr{R}_1 \equiv  \underset{m \in \textbf{N}}{\underset{\gamma \in \Delta_+ ( A)}{\prod}} \mathrm{exp}_q \bigg[ ( q - q^{-1} ) s^{-1}_{m,\gamma} e_{\gamma+m\delta} \otimes f_{\gamma + m \delta}  \bigg]   \text{, } \\ \mathscr{R}_2 \equiv    \mathrm{exp} \bigg[ ( q - q^{-1} ) \underset{m \in \textbf{Z}}{\sum} \text{ }  \overset{r}{\underset{i-j, i^{\prime}-j^{\prime}=1}{\sum}}                u_m ( i - j ) ( i^{\prime} - j^{\prime} ) e_{m\delta , \alpha_{i-j}}  \otimes f_{m\delta , \alpha_{i^{\prime}-j^{\prime}}}                    \bigg]  \text{, } \\ \mathscr{R}_3 \equiv      \underset{ m \in \textbf{N}}{\underset{\gamma \in \Delta_+ ( A)}{\prod}}     \mathrm{exp}_q \bigg[ \big( q - q^{-1} \big) s^{-1}_{m,\gamma-\delta} e_{\delta-\gamma+m \delta} \otimes f_{\delta-\gamma + m \delta}   \bigg]  \text{, } \\ \mathscr{R}_4 \equiv      \underset{ m \in \textbf{N}}{\underset{\gamma \in \Delta_+ ( A)}{\prod}}   \mathrm{exp} \bigg[ \hbar              \overset{r}{\underset{j\neq j^{\prime}}{\underset{i-j,i-j^{\prime}=1}{\sum}}}   \beta_{(i-j)(i-j^{\prime})} h_{\alpha(i-j)}   \otimes h_{\alpha(i-j^{\prime})}   \bigg]         \text{, }
\end{align*}

\noindent from the factorization of the Universal R-matrix to,

\begin{align*}
     \beta_l \big( u \big) \beta_{l+1} \big( v \big)     \text{, }
\end{align*}

\noindent in the first relation, (1), which appears in the set of relations for (*)(1). The RHS of the same relation has a prefactor of the form,

\begin{align*}
  W_{(1,1)}  \text{, }
\end{align*}

\noindent corresponding to the entry of the Boltzmann weight matrix which is applied to,

\begin{align*}
  \beta_l \big( v \big) \beta_{l+1} \big( u \big)   \text{, }
\end{align*}

\noindent which can be used to obtain the desired representation of each intertwining vector through the a series of manipulations originally introduced for the rational 7-vertex model for obtaining two-dimensional intertwining vectors in {\color{blue}[1]}. In the presence of additional degrees of freedom for the 20-vertex model, a generalization of the identities which hold when multiplying together terms of the R-matrix for the rational 7-vertex can be used for the universal R-matrix factorization of the 20-vertex model. Write,

\begin{align*}
     \big[ \mathcal{R}_{(1,1)} + \mathcal{R}_{(1,4)} \big]   \beta_l \big( u \big)  \beta_{l+1} \big( v \big)    \text{, }
\end{align*}

\noindent which can be alternatively expressed, with appropriate q-exponential terms from the factorization of the Universal R-matrix, with,

\begin{align*}
    \mathscr{T}_1 \mathscr{T}_2 \mathscr{T}_3 \mathscr{T}^{\prime}_3 +             \mathscr{T}_4   \mathscr{T}^{\prime}_4  \mathscr{T}_5 \mathscr{T}_6    \text{,}
\end{align*}

\noindent where,

\begin{align*}
     \mathscr{T}_1 \equiv   \underset{m \in \textbf{N}}{\underset{\gamma \in \Delta_+ ( A)}{\prod}} \mathrm{exp}_q \bigg[ ( q - q^{-1} ) s^{-1}_{m,\gamma} e_{\gamma+m\delta} \otimes f_{\gamma + m \delta}  \bigg] \text{, }  \\ \mathscr{T}_2 \equiv    \mathrm{exp} \bigg[ ( q - q^{-1} ) \underset{m \in \textbf{Z}}{\sum} \text{ }  \overset{r}{\underset{i-j, i^{\prime}-j^{\prime}=1}{\sum}}                u_m ( i - j ) ( i^{\prime} - j^{\prime} ) e_{m\delta , \alpha_{i-j}}  \otimes f_{m\delta , \alpha_{i^{\prime}-j^{\prime}}}                    \bigg] \text{, }  \\  \mathscr{T}_3 \equiv      \underset{ m \in \textbf{N}}{\underset{\gamma \in \Delta_+ ( A)}{\prod}}     \mathrm{exp}_q \bigg[ \big( q - q^{-1} \big) s^{-1}_{m,\gamma-\delta} e_{\delta-\gamma+m \delta} \otimes f_{\delta-\gamma + m \delta}   \bigg]  \text{, } \\ \mathscr{T}^{\prime}_3 \equiv      \underset{ m \in \textbf{N}}{\underset{\gamma \in \Delta_+ ( A)}{\prod}}    \mathrm{exp} \bigg[ \hbar              \overset{r}{\underset{j\neq j^{\prime}}{\underset{i-j,i-j^{\prime}=1}{\sum}}}   \beta_{(i-j)(i-j^{\prime})} h_{\alpha(i-j)}   \otimes h_{\alpha(i-j^{\prime})}   \bigg]    \text{, }     \\ \mathscr{T}_4 \equiv    \underset{m \in \textbf{N}}{\underset{\gamma \in \Delta_+ ( A)}{\prod}} \mathrm{exp}_q \bigg[ ( q - q^{-1} ) s^{-1}_{m,\gamma} e_{\gamma+m\delta} \otimes f_{\gamma + m \delta}  \bigg] \text{, } \\ \mathscr{T}^{\prime}_4 \equiv  \underset{m \in \textbf{N}}{\underset{\gamma \in \Delta_+ ( A)}{\prod}}   \mathrm{exp} \bigg[ ( q - q^{-1} ) \underset{m \in \textbf{Z}}{\sum} \text{ }  \overset{r}{\underset{i-j, i^{\prime}-j^{\prime}=4}{\sum}}                u_m ( i - j ) ( i^{\prime} - j^{\prime} ) e_{m\delta , \alpha_{i-j}}  \otimes f_{m\delta , \alpha_{i^{\prime}-j^{\prime}}}                    \bigg] \text{, } \\ \mathscr{T}_5 \equiv    \underset{ m \in \textbf{N}}{\underset{\gamma \in \Delta_+ ( A)}{\prod}}     \mathrm{exp}_q \bigg[ \big( q - q^{-1} \big) s^{-1}_{m,\gamma-\delta} e_{\delta-\gamma+m \delta} \otimes f_{\delta-\gamma + m \delta}   \bigg]  \text{, }    \\ \mathscr{T}_6 \equiv    \mathrm{exp} \bigg[ \hbar              \overset{r}{\underset{j\neq j^{\prime}}{\underset{i-j,i-j^{\prime}=4}{\sum}}}   \beta_{(i-j)(i-j^{\prime})} h_{\alpha(i-j)}   \otimes h_{\alpha(i-j^{\prime})}   \bigg]               \text{. }
\end{align*}

\noindent The product from the expressions above equals,

\begin{align*}
  \mathcal{T}_1 \mathcal{T}_2 \mathcal{T}_3 \mathcal{T}^{\prime}_3 + \mathcal{T}_4 \mathcal{T}^{\prime}_4    \text{, } 
\end{align*}

\noindent where,

\begin{align*}
   \mathcal{T}_1 \equiv   \underset{m \in \textbf{N}}{\underset{\gamma \in \Delta_+ ( A)}{\prod}} \mathrm{exp}_q \bigg[ ( q - q^{-1} ) s^{-1}_{m,\gamma} e_{\gamma+m\delta} \otimes f_{\gamma + m \delta}  \bigg]       \text{, }  \\ \mathcal{T}_2 \equiv    \underset{m \in \textbf{N}}{\underset{\gamma \in \Delta_+ ( A)}{\prod}}    \mathrm{exp} \bigg[ ( q - q^{-1} ) \underset{m \in \textbf{Z}}{\sum} \text{ }  \overset{r}{\underset{i-j, i^{\prime}-j^{\prime}=4}{\sum}}                u_m ( i - j ) ( i^{\prime} - j^{\prime} ) e_{m\delta , \alpha_{i-j}}   \otimes f_{m\delta , \alpha_{i^{\prime}-j^{\prime}}}                    \bigg] \text{, } \\  \mathcal{T}_3 \equiv     
     \mathrm{exp} \bigg[ ( q - q^{-1} ) \underset{m \in \textbf{Z}}{\sum} \text{ }  \overset{4}{\underset{i-j, i^{\prime}-j^{\prime}=1}{\sum}}                u_m ( i - j ) ( i^{\prime} - j^{\prime} ) e_{m\delta , \alpha_{i-j}}   \otimes f_{m\delta , \alpha_{i^{\prime}-j^{\prime}}}                    \bigg] \text{, } \\ \mathcal{T}^{\prime}_3 \equiv   \mathrm{exp} \bigg[ \hbar              \overset{4}{\underset{j\neq j^{\prime}}{\underset{i-j,i-j^{\prime}=1}{\sum}}}   \beta_{(i-j)(i-j^{\prime})} h_{\alpha(i-j)}   \otimes h_{\alpha(i-j^{\prime})}   \bigg]               + 1 \text{, }  \\           \mathcal{T}_4 \equiv    \underset{ m \in \textbf{N}}{\underset{\gamma \in \Delta_+ ( A)}{\prod}}     \mathrm{exp}_q \bigg[ \big( q - q^{-1} \big) s^{-1}_{m,\gamma-\delta} e_{\delta-\gamma+m \delta}  \otimes f_{\delta-\gamma + m \delta}   \bigg]  \text{, } \\     \mathcal{T}^{\prime}_4 \equiv   \mathrm{exp} \bigg[ \hbar              \overset{r}{\underset{j\neq j^{\prime}}{\underset{i-j,i-j^{\prime}=4}{\sum}}}   \beta_{(i-j)(i-j^{\prime})} h_{\alpha(i-j)}   \otimes h_{\alpha(i-j^{\prime})}              \bigg]     \text{, }
\end{align*}

\noindent upon collecting like terms. As a generalization of the identities that hold over two-dimensions for the rational 7-vertex model, the expression above,

\begin{align*}
     \big[ \mathcal{R}_{(1,1)} + \mathcal{R}_{(1,4)} \big]   \beta_l \big( u \big)  \beta_{l+1} \big( v \big)    \text{, }
\end{align*}

\noindent in presence of the factorization of bulk and boundary conditions for the universal R-matrix, equals,

\begin{align*}
        \big[ \mathcal{R}_{(1,2)} + \mathcal{R}_{(1,3)} \big] \beta_l \big( u \big) \beta_{l+1} \big( v \big)          \text{, }
\end{align*}

\noindent which in the same manner as demonstrated above for the q-exponential factors appearing in,

\begin{align*}
  \mathcal{R}_{(1,1)} + \mathcal{R}_{(1,4)}  \text{, }
\end{align*}

\noindent for,

\begin{align*}
  \mathcal{R}_{(1,2)} + \mathcal{R}_{(1,3)}  \text{, }
\end{align*}

\noindent can be expressed with,

\begin{align*}
  \mathcal{R}_1 \mathcal{R}_2 \mathcal{R}_3 \mathcal{R}_4 \mathcal{R}_5 \mathcal{R}_6   \text{, }
\end{align*}

\noindent where, 

\begin{align*}
  \mathcal{R}_1 \equiv    \underset{m \in \textbf{N}}{\underset{\gamma \in \Delta_+ ( A)}{\prod}} \mathrm{exp}_q \bigg[ ( q - q^{-1} ) s^{-1}_{m,\gamma} e_{\gamma+m\delta} \otimes f_{\gamma + m \delta}  \bigg]       \text{, }   \\ \mathcal{R}_2 \equiv  \underset{m \in \textbf{N}}{\underset{\gamma \in \Delta_+ ( A)}{\prod}}     \mathrm{exp} \bigg[ ( q - q^{-1} ) \underset{m \in \textbf{Z}}{\sum} \text{ }  \overset{r}{\underset{i-j, i^{\prime}-j^{\prime}=2}{\sum}}                u_m ( i - j ) ( i^{\prime} - j^{\prime} ) e_{m\delta , \alpha_{i-j}}   \otimes f_{m\delta , \alpha_{i^{\prime}-j^{\prime}}}                    \bigg]     \\  \mathcal{R}_3 \equiv  \mathrm{exp} \bigg[ ( q - q^{-1} ) \underset{m \in \textbf{Z}}{\sum} \text{ }  \overset{3}{\underset{i-j, i^{\prime}-j^{\prime}=2}{\sum}}                u_m ( i - j ) ( i^{\prime} - j^{\prime} ) e_{m\delta , \alpha_{i-j}}  \otimes f_{m\delta , \alpha_{i^{\prime}-j^{\prime}}}                    \bigg] \text{, }\\ \mathcal{R}_4 \equiv  \mathrm{exp} \bigg[ \hbar              \overset{3}{\underset{j\neq j^{\prime}}{\underset{i-j,i-j^{\prime}=2}{\sum}}}   \beta_{(i-j)(i-j^{\prime})} h_{\alpha(i-j)}    \otimes h_{\alpha(i-j^{\prime})}   \bigg]                  + 1  \text{, } \\  \mathcal{R}_5 \equiv     \underset{ m \in \textbf{N}}{\underset{\gamma \in \Delta_+ ( A)}{\prod}}     \mathrm{exp}_q \bigg[ \big( q - q^{-1} \big) s^{-1}_{m,\gamma-\delta} e_{\delta-\gamma+m \delta}   \otimes f_{\delta-\gamma + m \delta}   \bigg]   \text{, } \\ \mathcal{R}_6 \equiv   \mathrm{exp} \bigg[ \hbar              \overset{r}{\underset{j\neq j^{\prime}}{\underset{i-j,i-j^{\prime}=2}{\sum}}}   \beta_{(i-j)(i-j^{\prime})} h_{\alpha(i-j)}   \otimes h_{\alpha(i-j^{\prime})}   \bigg]        \text{, }
\end{align*}

\noindent upon collecting like terms. To isolate the action of the product $\beta_l \big( u \big) \beta_{l+1} \big( v \big)$ of intertwining vectors, it suffices to more closely examine an expression of the form,

\begin{align*}
  \frac{\underline{\mathcal{T}_1 }}{\underline{\mathcal{T}_q}}  \underline{\mathcal{T}_q} \bigg[  \frac{\underline{\mathcal{T}_2} \underline{\mathcal{T}_3} +1}{\underline{\mathcal{T}^{\prime}_2} \underline{\mathcal{T}^{\prime}_3} + 1} \bigg]  \frac{\underline{\mathcal{T}_4}}{\underline{\mathcal{T}_q}}  \underline{\mathcal{T}_q}     \equiv  \underline{\mathcal{T}_1} \bigg[  \frac{\underline{\mathcal{T}_2} \underline{\mathcal{T}_3} +1}{\underline{\mathcal{T}^{\prime}_2} \underline{\mathcal{T}^{\prime}_3} + 1} \bigg]  \underline{\mathcal{T}_4}  \text{, }
\end{align*}

\noindent where,

\begin{align*}
  \underline{\mathcal{T}_1} \equiv   \mathrm{exp} \bigg[ \big( q - q^{-1} \big) \underset{m \in \textbf{Z}}{\sum} \text{ } \overset{4}{\underset{i-j, i^{\prime} - j^{\prime}=2}{\sum}}    u_m \big( i - j \big) \big( i^{\prime} - j^{\prime} \big) e_{m\delta , \alpha_{i-j}} \otimes f_{m \delta, \alpha_{i^{\prime}-j^{\prime}}}       \bigg]       \text{, } \\ \underline{\mathcal{T}_2} \equiv   \mathrm{exp} \bigg[ ( q - q^{-1} ) \underset{m \in \textbf{Z}}{\sum} \text{ }  \overset{4}{\underset{i-j, i^{\prime}-j^{\prime}=1}{\sum}}                u_m ( i - j ) ( i^{\prime} - j^{\prime} ) e_{m\delta , \alpha_{i-j}}  \otimes f_{m\delta , \alpha_{i^{\prime}-j^{\prime}}}                    \bigg]                    \text{, } \end{align*}

  \begin{align*}
  \underline{\mathcal{T}_3} \equiv   \mathrm{exp} \bigg[ \hbar              \overset{4}{\underset{j\neq j^{\prime}}{\underset{i-j,i-j^{\prime}=1}{\sum}}}   \beta_{(i-j)(i-j^{\prime})} h_{\alpha(i-j)}   \otimes h_{\alpha(i-j^{\prime})}   \bigg]             \text{, } \\         
 \underline{\mathcal{T}^{\prime}_2} \equiv \mathrm{exp} \bigg[ ( q - q^{-1} ) \underset{m \in \textbf{Z}}{\sum} \text{ }  \overset{2}{\underset{i-j, i^{\prime}-j^{\prime}=1}{\sum}}                u_m ( i - j ) ( i^{\prime} - j^{\prime} ) e_{m\delta , \alpha_{i-j}}  \otimes f_{m\delta , \alpha_{i^{\prime}-j^{\prime}}}                    \bigg]  \text{, } \\  
    \underline{\mathcal{T}^{\prime}_3} \equiv     \mathrm{exp} \bigg[ \hbar              \overset{3}{\underset{j\neq j^{\prime}}{\underset{i-j,i-j^{\prime}=2}{\sum}}}   \beta_{(i-j)(i-j^{\prime})} h_{\alpha(i-j)}   \otimes h_{\alpha(i-j^{\prime})}   \bigg]               \text{, } \\   \underline{\mathcal{T}_4} \equiv    \mathrm{exp} \bigg[ \hbar              \overset{4}{\underset{j\neq j^{\prime}}{\underset{i-j,i-j^{\prime}=2}{\sum}}}   \beta_{(i-j)(i-j^{\prime})} h_{\alpha(i-j)}   \otimes h_{\alpha(i-j^{\prime})}   \bigg]                  \text{, }
\end{align*}

\noindent and the cancellation factor,

\begin{align*}
 \underline{\mathcal{T}_q} \equiv        \underset{m \in \textbf{N}}{\underset{\gamma \in \Delta_+ ( A)}{\prod}} \mathrm{exp}_q \bigg[ ( q - q^{-1} ) s^{-1}_{m,\gamma} e_{\gamma+m\delta} \otimes f_{\gamma + m \delta}  \bigg]                 \text{. }
\end{align*}

\noindent from the observation that the first term, and the last term, $\mathcal{T}_1$ and $\mathcal{T}_4$, respectively, are obtained from subtracting,

\begin{align*}
   \underset{m \in \textbf{Z}}{\sum} \bigg[  \bigg[   \overset{r}{\underset{i-j,i^{\prime}-j^{\prime}=4}{\sum}}  - \overset{r}{\underset{i-j,i^{\prime}-j^{\prime}=2}{\sum}}    \bigg]           u_m ( i - j ) ( i^{\prime} - j^{\prime} ) e_{m\delta , \alpha_{i-j}}  \otimes f_{m\delta , \alpha_{i^{\prime}-j^{\prime}}}   \bigg]               
 \text{, }
\end{align*}

\noindent and,

\begin{align*}
  \hbar \bigg[ \bigg[ \underset{j \neq j^{\prime}}{\underset{i-j,i-j^{\prime}=4}{\sum}}  - \underset{j \neq j^{\prime}}{\underset{i-j,i-j^{\prime}=2}{\sum}}  \bigg]      \beta_{(i-j)(i-j^{\prime})} h_{\alpha(i-j)}   \otimes h_{\alpha(i-j^{\prime})}       \bigg]  \text{, }
\end{align*}

\noindent in the power of the q-exponential, and exponential. Making use of these observations implies that the 20-vertex intertwining vector can be obtained from solutions to,

\begin{align*}
 \bigg[   \underline{\mathcal{T}_1} \bigg[  \frac{\underline{\mathcal{T}_2} \underline{\mathcal{T}_3} +1}{\underline{\mathcal{T}^{\prime}_2} \underline{\mathcal{T}^{\prime}_3} + 1} \bigg]  \underline{\mathcal{T}_4}   - 1 \bigg]  \beta_l \big( v \big) \beta_{l+1} \big( u \big) = 0  \text{, }
\end{align*}

\noindent which can be rearranged into the following series of equivalent relations,

\begin{align*}
       \frac{\beta_l \big( v \big)}{\beta_{l+1} \big( u \big) }         =  \bigg[  \underline{\mathcal{T}_1} \bigg[  \frac{\underline{\mathcal{T}_2} \underline{\mathcal{T}_3} +1}{\underline{\mathcal{T}^{\prime}_2} \underline{\mathcal{T}^{\prime}_3} + 1} \bigg]  \underline{\mathcal{T}_4}  \bigg]   \frac{\beta_l \big( v \big)}{\beta_{l+1} \big( u \big) }   \\ \Updownarrow  \\    \frac{ \gamma_l \big( u \big) \beta_{l+1} \big( v \big)}{\beta_l  \big( u \big) \gamma_{l+1} \big( v \big)  }      =  \bigg[  \underline{\mathcal{T}_1} \bigg[  \frac{\underline{\mathcal{T}_2} \underline{\mathcal{T}_3} +1}{\underline{\mathcal{T}^{\prime}_2} \underline{\mathcal{T}^{\prime}_3} + 1} \bigg]  \underline{\mathcal{T}_4}  \bigg]   \frac{ \gamma_l \big( u \big) \beta_{l+1} \big( v \big)}{\beta_l  \big( u \big) \gamma_{l+1} \big( v \big)  }    \\ \Updownarrow  \\                    \frac{ Z_l \big( u \big) \beta_{l+1} \big( v \big)}{\beta_l  \big( u \big) Z_{l+1} \big( v \big)  }      =  \bigg[  \underline{\mathcal{T}_1} \bigg[  \frac{\underline{\mathcal{T}_2} \underline{\mathcal{T}_3} +1}{\underline{\mathcal{T}^{\prime}_2} \underline{\mathcal{T}^{\prime}_3} + 1} \bigg]  \underline{\mathcal{T}_4}  \bigg]   \frac{ Z_l \big( u \big) \beta_{l+1} \big( v \big)}{\beta_l  \big( u \big) Z_{l+1} \big( v \big)  }   \\ \Updownarrow  \\                     \frac{ Z_l \big( u \big) \gamma_{l+1} \big( v \big)}{\gamma_l  \big( u \big) Z_{l+1} \big( v \big)  }      =  \bigg[  \underline{\mathcal{T}_1} \bigg[  \frac{\underline{\mathcal{T}_2} \underline{\mathcal{T}_3} +1}{\underline{\mathcal{T}^{\prime}_2} \underline{\mathcal{T}^{\prime}_3} + 1} \bigg]  \underline{\mathcal{T}_4}  \bigg]   \frac{ Z_l \big( u \big) \gamma_{l+1} \big( v \big)}{\gamma_l  \big( u \big) Z_{l+1} \big( v \big)  }  \\  \Updownarrow \\                    \frac{ \beta_l \big( u \big) \gamma_{l+1} \big( v \big)}{\gamma_l  \big( u \big) \beta_{l+1} \big( v \big)  }      =  \bigg[  \underline{\mathcal{T}_1} \bigg[  \frac{\underline{\mathcal{T}_2} \underline{\mathcal{T}_3} +1}{\underline{\mathcal{T}^{\prime}_2} \underline{\mathcal{T}^{\prime}_3} + 1} \bigg]  \underline{\mathcal{T}_4}  \bigg]   \frac{ \beta_l \big( u \big) \gamma_{l+1} \big( v \big)}{\beta_l  \big( u \big) Z_{l+1} \big( v \big)  }  
 \text{, }
\end{align*}

\noindent by multiplying, and dividing, by,

\begin{align*}
  \gamma_l \big( u \big) \beta_{l+1} \big( v \big)   \text{, } \\ Z_l \big( u \big) \beta_{l+1} \big( v \big)    \text{, } \end{align*}

  \begin{align*}
  Z_l \big( u \big) \gamma_{l+1} \big( v \big)  \text{, } \\  \beta_l \big( u \big) \gamma_{l+1} \big( v \big)  \text{, }
\end{align*}

\noindent respectively. Any one of the equations above can be used to determine the components of the three intertwining vectors, from the observation that, instead of a product of two intertwining vectors from the SOS 20-vertex model, a product of three intertwining vectors take on the form below. For a third spectral parameter $\underline{w} \equiv w$,

\begin{align*}
       \frac{\beta_l \big( v \big)}{\beta_{l+1} \big( u \big) }         =  \bigg[  \underline{\mathcal{T}_1} \bigg[  \frac{\underline{\mathcal{T}_2} \underline{\mathcal{T}_3} +1}{\underline{\mathcal{T}^{\prime}_2} \underline{\mathcal{T}^{\prime}_3} + 1} \bigg]  \underline{\mathcal{T}_4}  \bigg]   \frac{\beta_l \big( v \big)}{\beta_{l+1} \big( u \big) }   \\ \Updownarrow \\     \frac{ \gamma_l \big( u \big) \beta_{l+1} \big( v \big) Z_{l+1} \big( w \big) }{\beta_l  \big( u \big) \gamma_{l+1} \big( v \big) Z_{l+1} \big( w \big) }      =  \bigg[  \underline{\mathcal{T}_1} \bigg[  \frac{\underline{\mathcal{T}_2} \underline{\mathcal{T}_3} +1}{\underline{\mathcal{T}^{\prime}_2} \underline{\mathcal{T}^{\prime}_3} + 1} \bigg]  \underline{\mathcal{T}_4}  \bigg]   \frac{ \gamma_l \big( u \big) \beta_{l+1} \big( v \big)  }{\beta_l  \big( u \big) \gamma_{l+1} \big( v \big)    }   \frac{Z_{l+1} \big( w \big)}{ Z_{l+1} \big( w \big)}     \\ \Updownarrow \\                    \frac{ Z_l \big( u \big) \beta_{l+1} \big( v \big) \gamma_{l+2} \big( w \big) }{\beta_l  \big( u \big) Z_{l+1} \big( v \big) \gamma_{l+2} \big( w \big)  }      =  \bigg[  \underline{\mathcal{T}_1} \bigg[  \frac{\underline{\mathcal{T}_2} \underline{\mathcal{T}_3} +1}{\underline{\mathcal{T}^{\prime}_2} \underline{\mathcal{T}^{\prime}_3} + 1} \bigg]  \underline{\mathcal{T}_4}  \bigg]   \frac{ Z_l \big( u \big) \beta_{l+1} \big( v \big) \gamma_{l+2} \big( w \big) }{\beta_l  \big( u \big) Z_{l+1} \big( v \big)  \gamma_{l+2} \big( w \big) }   \\ \Updownarrow \\                    \frac{ Z_l \big( u \big) \gamma_{l+1} \big( v \big) \beta_{l+2} \big( w \big) }{\gamma_l  \big( u \big) Z_{l+1} \big( v \big) \beta_{l+2} \big( w \big)  }      =  \bigg[  \underline{\mathcal{T}_1} \bigg[  \frac{\underline{\mathcal{T}_2} \underline{\mathcal{T}_3} +1}{\underline{\mathcal{T}^{\prime}_2} \underline{\mathcal{T}^{\prime}_3} + 1} \bigg]  \underline{\mathcal{T}_4}  \bigg]   \frac{ Z_l \big( u \big) \gamma_{l+1} \big( v \big) \gamma_{l+2} \big( w \big)  }{\gamma_l  \big( u \big) Z_{l+1} \big( v \big) \gamma_{l+2} \big( w \big)  }  \\  \Updownarrow \\                    \frac{ \beta_l \big( u \big) \gamma_{l+1} \big( v \big) Z_{l+2} \big( w \big) }{\gamma_l  \big( u \big) \beta_{l+1} \big( v \big) Z_{l+2} \big( w \big)  }      =  \bigg[  \underline{\mathcal{T}_1} \bigg[  \frac{\underline{\mathcal{T}_2} \underline{\mathcal{T}_3} +1}{\underline{\mathcal{T}^{\prime}_2} \underline{\mathcal{T}^{\prime}_3} + 1} \bigg]  \underline{\mathcal{T}_4}  \bigg]   \frac{ \beta_l \big( u \big) \gamma_{l+1} \big( v \big)  Z_{l+2} \big( w \big)   }{\beta_l  \big( u \big) Z_{l+1} \big( v \big) Z_{l+2} \big( w \big)  }  
 \text{, }
\end{align*}

\noindent by multiplying, and dividing, by,

\begin{align*}
  \gamma_l \big( u \big) \beta_{l+1} \big( v \big)  Z_{l+2} \big( w \big)    \text{, } \\ Z_l \big( u \big) \beta_{l+1} \big( v \big)  \gamma_{l+2} \big( w \big)   \text{, } \\ 
  Z_l \big( u \big) \gamma_{l+1} \big( v \big)  \beta_{l+2} \big( w \big)  \text{, } \\  \beta_l \big( u \big) \gamma_{l+1} \big( v \big)  Z_{l+2} \big( w \big)  \text{, }
\end{align*}

\noindent respectively. WLOG setting the first two components of the first two intertwining vectors,

\begin{align*}
     \gamma_l \big( u \big)   \text{, } \\    \beta_l \big( u \big)   \text{, }  
\end{align*}

\noindent to $1$,

\begin{align*}
     \gamma_l \big( u \big) \equiv \bigg[ \begin{smallmatrix} 1 \\ 1 \\ *_{\gamma} \end{smallmatrix}  \bigg]  \text{, } \\    \beta_l \big( u \big) \equiv \bigg[ \begin{smallmatrix} 1 \\ 1 \\ *_{\beta} \end{smallmatrix} \bigg]   \text{, }  
\end{align*} 

\noindent implies that the remaining third entry of the first two inertwinning vectors, $*_{\gamma}$ and $*_{\beta}$, above each take the form,

\begin{align*}
    *_{\gamma} \equiv  \gamma_l \big( u_3 \big)  \equiv \underline{\mathcal{T}_1} \bigg[  \frac{\underline{\mathcal{T}_2} \underline{\mathcal{T}_3} +1}{\underline{\mathcal{T}^{\prime}_2} \underline{\mathcal{T}^{\prime}_3} + 1}   \bigg]  \text{, } \\ *_{\beta}  \equiv  \beta_l \big( u_3 \big) \equiv \bigg[  \underline{\mathcal{T}_1} \bigg[  \frac{\underline{\mathcal{T}_2} \underline{\mathcal{T}_3} +1}{\underline{\mathcal{T}^{\prime}_2} \underline{\mathcal{T}^{\prime}_3} + 1} \bigg] \bigg]^{-1}    \text{, }
\end{align*}

\noindent respectively, from the following two relations, the first of which has the prefactor,

\begin{align*}
 \frac{Z_l \big( u \big) \gamma_{l+1} \big(  v \big) \beta_{l+2} \big( w \big) }{\gamma_l \big( u \big) Z_{l+1} \big( v \big)  \beta_{l+2} \big( w \big) }    \text{, }
\end{align*}

\noindent which after applying the intertwining relation, implies that the relations above can be used to obtain the Solid-on-Solid intertwining vectors for the 20-vertex model, namely,

\begin{align*}
 \psi^{20V} \big( u \big)^a_{b^{\prime}}   \text{. }
\end{align*}

\noindent WLOG, the second intertwining vector included in the set of linear combinations,

\begin{align*}
  \underset{\underline{u} \equiv u , \underline{v} \equiv v , \in V ( \textbf{T})}{\underset{\textbf{T}}{\mathrm{span} }} \big\{  \psi^{20V} \big( u \big)^a_{b^{\prime}},  \psi^{20V} \big( u \big)^{b^{\prime}}_c  \big\}  \text{, }
\end{align*}

\noindent takes the desired form after setting the third, instead of the second, component of the intertwining vector to equal $1$. As a result, the final remaining entr - the second entry - of the second intertwining vector equals the reciprocal of the constant which appeared in the third entry of the first inertwining vector. The desired set of expressions for the Boltzmann weight matrix of the Solid-on-Solid 20-vertex model can be obtained from the observation that, 

\begin{align*}
    W \bigg[   \begin{smallmatrix}
     l + 2  &  l+1   \\ l + 1  & l + 1  \\  l & l + 1
  \end{smallmatrix}\bigg| \underline{u} \bigg]  \text{, }
\end{align*}

\noindent which corresponds to the first possible value of the Boltzmann weight matrix, equals,

\begin{align*}
   \mathcal{R}_{(1,1)} \big( u \big)  + 1 = \mathcal{R}_{(1,1)} + 1   \equiv \mathcal{C}_1  \text{. } 
\end{align*}

\noindent For the remaining possible values of the Boltzmmann weight matrix, each one of the entries equals,

\begin{align*}
     \frac{\mathcal{R}_{(1,1)}  \big( u + l + w  \big)}{\mathcal{R}_{(1,1)}  \big( l + w  \big)}     \equiv \mathcal{C}_2     \text{, } \\           \frac{\mathcal{R}_{(1,1)}  \big( u  \big( 1 + l + w \big)   \big)}{\mathcal{R}_{(1,1)}  \big( l + w  \big)}       \equiv \mathcal{C}_3         \text{, }
\end{align*}

\noindent respectively, for,

\begin{align*}
    l \equiv  \bigg[ \begin{smallmatrix} \textbf{R} \\ \textbf{R} \end{smallmatrix} \bigg]  \text{, } \\   w \equiv \bigg[  \begin{smallmatrix} \textbf{R} \\ \textbf{R} \end{smallmatrix}  \bigg] \text{, }
\end{align*}

\noindent from which we conclude the argument. \boxed{ }

\section{Conclusion}

\noindent In this paper, to establish several points of comparison between vertex, and Solid-on-Solid, models, we draw the attention of the reader to a previous approach, due to Antoenneko and Valinevich, on the 7-vertex model. As a two-dimensional vertex model, the 7-vertex model is of great interest not only in the similarities that it raises with the 6-vertex model, particularly from the fact that its sample space has an additional configuration within the probability ensemble, but also from the fact that the construction for the R-matrix, and the resulting Bolzmann weight matrix for the Solid-on-Solid model, can be conceptualized for three-dimensional vertex models. For the 20-vertex model, from leveraging previous work of the author through an adaptation of the seminal QISM framework, arguments in previous section of the paper demonstrated how one can: (1) incorporate Integrable Probabilistic objects with Discrete Probabilistic objects through transfer, and quantum monodromy, matrices, and action-angle coordinates; (2) formulate a system of equations, for determining entries of the three-dimensional Boltzmann weight matrix, from a factorization of the universal R-matrix; (3) obtain explicit formulas for entries of three-dimensional intertwining vectors. Future applications, and adaptations, of Integrability, vertex models, whether two, or three, dimensional, and Solid-on-Solid models, are of interest to further explore.

\section*{Declarations}

\subsection{Ethics approval and consent to participate}

The author consents to participate in the peer review process.

\subsection{Consent for publication}

The author consents to submit the following work for publication.

\subsection{Availability of data and materials}

Not applicable

\subsection{Competing interests}

Not applicable

\subsection{Funding}

Not applicable

\nocite{*}
\bibliography{sn-bibliography}

\end{document}